\documentclass[a4,10pt, usenatbib]{mn2e}
\usepackage[dvips]{graphicx, psfrag}
\usepackage{times}
\usepackage{amsmath}

\usepackage{natbib}
\usepackage{aas_macros}

\title[Key to large toroidal
magnetic fields within stars]
{Coexistence of oppositely flowing 
multi-$\varphi$-currents: Key to large toroidal 
magnetic fields within stars}
\author[K. Fujisawa \& Y. Eriguchi]{Kotaro
Fujisawa
\thanks{E-mail: fujisawa@ea.c.u-tokyo.ac.jp}
and Yoshiharu Eriguchi \\
Department of Earth Science and Astronomy,
Graduate School of Arts and Sciences, University of Tokyo,\\
Komaba, Meguro-ku, Tokyo 153-8902, Japan}

\date{Accepted 2013 Mar 25. Received 2013 Mar 23; in original form 2013 February 5}

\def\Vec#1{\mbox{\boldmath $#1$}}
\def\D#1#2{\dfrac{d #1}{d #2}}

\def\P#1#2{\dfrac{\partial #1}{\partial #2}}
\def\PP#1#2{\dfrac{\partial^2 #1}{\partial #2^2}}

\begin{document}

\maketitle

\begin{abstract}

We will show the importance of coexistence
of oppositely flowing $\varphi$-currents
for magnetized stars to sustain 
strong toroidal magnetic fields  
within the stars by analyzing stationary 
states of magnetized stars with surface 
currents which flow in the opposite direction
with respect to the bulk currents within the stars.
We have imposed boundary 
conditions for currents and 
toroidal magnetic fields to vanish outside the stars.
It is important to note that these boundary 
conditions set an upper limit for the total current 
within the stars. This upper limit for the
total current results in the presence of an
upper limit for the magnitude of the energy 
for the toroidal magnetic fields of the stars.  
If the stars could have the toroidal 
surface currents which flow in the opposite 
directions to the internal toroidal currents, the 
positively flowing internal toroidal currents 
can become stronger than the upper limit value 
of the current for
configurations without surface toroidal currents.
Thus the energies for the toroidal magnetic fields 
can become much larger than those for the
magnetized stars without surface toroidal 
currents. We have also analyzed the same 
phenomena appearing in spherical incompressible 
stars for dipole-like magnetic fields with or without
surface toroidal currents by employing the
zero-flux-boundary method. 
We have applied those configurations
with surface toroidal currents to magnetars
and discussed their flares through
which magnetic helicities could arise 
outside the stellar surfaces.

\end{abstract}

\section{Introduction}

The origin of the magnetic fields inside the star is 
not still understood well. According to the recent 
progresses of the theoretical researches, there are 
two possibilities for the mechanism to generate and 
sustain magnetic fields, i.e. 1) the dynamo theory 
and 2) the fossil field theory (see the review by
\citealt{Moss_IAU162}). Stars which have convective 
regions regenerate and sustain the magnetic fields by 
the dynamo mechanism.  On the other hand, 
A, B, O stars have a convective core
where dynamo action develops
(e.g. \citealt{Brun_Browning_Toomre_2005} ).
However, the timescale for the dynamo-generated 
field in this core to reach the surface is 
too long (\citealt{Charbonneau_MacGregor_2001}).
Therefore the magnetic field observed at the 
stellar surface is supposed to be fossil field.
Recent observations show the observed 
magnetic fields' properties confirm this hypothesis 
(see \citealt{Wade_et_al_IAU272}).
Consequently, 
it has long been considered that magnetic fields 
of such stars would come from fossil magnetic fields. 
According to the fossil field theory, the magnetic 
fields of such stars would be originated from 
magnetized interstellar media. If the magnetic 
fluxes would be conserved during star formation 
processes, the magnetic fields could be 
concentrated to smaller regions of the stars 
by the gravitational contraction and result in 
strong magnetic fields inside the stars. 
Since the electric conductivities of the stars are 
very large, those magnetic fields do not diffuse 
during their formation stages. As for neutron stars, 
the origin of their strong magnetic fields is much 
more uncertain. If the fossil field theory could
be applied, the strong magnetic fields could be
reached by the same mechanism as those mentioned 
above for stars without external convective regions.
On the other hand, if we could adopt the dynamo theory, 
the strong magnetic fields would be formed  
by the dynamo due to the rapid differential 
rotations in the convective regions
of the proto-neutron stars (\citealt{Duncan_Thompson_1992}).
The differential rotations could wind up the 
initial poloidal magnetic fields to produce 
strong toroidal magnetic fields before the crusts 
would crystallize. In either case, the magnetic 
fields of neutron stars would be present even 
at birth and survive in much longer timescales
than the Alfv\'en timescale ($= \sqrt{4\pi 
\rho}r_s / B \sim 100$ s for typical neutron stars 
with magnetic fields of order of $B=10^{12}$G). 
Here $r_s$ is the radius of the stellar surface
which is, for axisymmetric configurations, a 
function of $\theta$ of the polar coordinates $(r, \theta, \varphi)$,
i.e. $r_s(\theta)$. In order to sustain these kinds of fossil magnetic 
fields for a long time, these magnetic fields 
must be stationary and stable. Thus,
in order to understand the magnetic fields
originated from the fossil fields, it would
be useful and important to get stationary and 
stable configurations of magnetized stars. 

As for stability, analytic studies have 
shown that any configurations with either 
purely poloidal or purely toroidal magnetic 
fields are unstable (\citealt{Tayler_1973}; 
\citealt{Markey_Tayler_1973}).
Stable magnetized stars should have both the 
poloidal and the toroidal 
magnetic fields. Moreover, the magnitudes of 
the toroidal fields must be comparable with 
those of the poloidal fields 
(\citealt{Tayler_1980}). This argument 
has been shown to be the case from the
recent simulations.
\cite{Braithwaite_Spruit_2004} have shown 
that an initial random magnetic field in 
stably stratified stellar layers 
relaxes on the stable twisted-torus magnetic field
configuration after several Alfv\'en timescale. 
Similar twisted-torus magnetic field 
configurations have been also obtained
in many previous works
by numerically exact computations of
the axisymmetric stationary states
of magnetized stars (\citealt{Tomimura_Eriguchi_2005}; 
\citealt{Yoshida_Eriguchi_2006}; \citealt{Lander_Jones_2009};
\citealt{Lander_Andersson_Glampedakis_2011}; 
\citealt{Fujisawa_Yoshida_Eriguchi_2012}), 
structure separated Grad-Shafranov (GS) solving method
(\citealt{Ciolfi_et_al_2009}; \citealt{Glampedakis_Andersson_Lander_2012})
or zero-flux-boundary method (\citealt{Prendergast_1956};
\citealt{Ioka_Sasaki_2004}; \citealt{Duez_Mathis_2010}; 
\citealt{Yoshida_Kiuchi_Shibata_2012}).
The stabilities of these fields, however, 
have not been clarified yet because 
it is difficult to analyze their
stability by linear stability analyses or
other means based on the stationary
configurations. On the other hand, \cite{Braithwaite_2009} 
and \cite{Duez_Braithwaite_Mathis_2010} 
have shown that the stability criteria of 
the magnetized stars could be 
expressed  as below:
\begin{eqnarray}
a \frac{{\cal M}}{|W|} < \frac{{\cal M}_p}{\cal M} \leq 0.8,
\label{Eq:criteria}
\end{eqnarray}
where ${\cal M} / {|W|}$ is the ratio of the 
magnetic energy to the gravitational energy 
(see Appendix \ref{App:Physics}). 
${\cal M}_p / {\cal M}$ is the ratio of the 
poloidal magnetic energy to the total magnetic 
energy and $a$ is a 
certain dimensionless factor of order 10 for 
main-sequence stars 
and of order $10^3$ for neutron stars. 
The value of ${\cal M}/{|W|}$ is about 
$10^{-6}$ even for magnetars
and is expected to be ${\cal M}/{|W|} < 10^{-6}$ 
for other real stars. Thus the left hand side of this 
inequality could be less than about $10^{-3}$
even if the value of $a$ might be $\sim 1000$. 
Therefore, this criterion means the
configurations with the twisted torus magnetic 
fields are stable even if the toroidal magnetic 
fields are much stronger than the poloidal 
magnetic fields. In contrast, the right hand 
side of this inequality means 
that the strong poloidal magnetic field configurations 
are unstable (we define that the poloidal fields 
are strong when ${\cal M}_p / {\cal M} > 0.8$ in 
this paper). As shown in dynamical simulations
mentioned above, configurations with the strong 
poloidal magnetic fields are likely to become
unstable within several  Alfv\'en timescales.
This criteria would not be applied to all 
situations because we might be able to
consider various kinds of magnetic field 
configurations as the initial states 
and different choices of the initial 
conditions might influence on the evolutions of 
the magnetic fields.
However, it seems to be the case that there is
a tendency to become more unstable even for
the twisted-torus magnetic field configurations 
with larger poloidal magnetic field energies. 
Therefore, it would be  a natural 
consequence to consider that there 
would be stable magnetized stars with
strong toroidal magnetic fields which satisfy
the condition ${\cal M}_p / {\cal M} < 0.8$, 
i.e. ${\cal M}_t / {\cal M} > 0.2$ where
${\cal M}_t$ is the energy of the toroidal field. 

On the other hand, the majority of investigations
in which stationary states of the magnetized stars 
have been treated (e.g. \citealt{Yoshida_Eriguchi_2006}; 
\citealt{Yoshida_Yoshida_Eriguchi_2006};
\citealt{Lander_Jones_2009}; \citealt{Ciolfi_et_al_2009}; \citealt{Lander_Andersson_Glampedakis_2011};
  \citealt{Fujisawa_Yoshida_Eriguchi_2012}) failed 
to obtain configurations with strong toroidal 
magnetic fields. In these studies,
stationary states of magnetized stars have
been pursued either by numerically
exact methods (e.g. \citealt{Tomimura_Eriguchi_2005}) or 
structure separated GS solving methods (e.g. \citealt{Ciolfi_et_al_2009}). However, 
they have only found that it was very difficult
to obtain stationary states of magnetized stars
with very strong toroidal magnetic fields.
In some of their solutions the toroidal 
magnetic fields have been almost as strong as 
the poloidal magnetic fields only in the particular 
local regions inside the stars, but the 
total energies of the  toroidal magnetic 
fields as a whole are much smaller than 
those of the total poloidal magnetic fields.
In other words, the ratios of 
${\cal M}_p / {\cal M}$ in their 
solutions are much bigger than 0.8.

By contrast, some studies of
magnetized stationary configurations
by structure separated GS solving method
(\citealt{Glampedakis_Andersson_Lander_2012}) 
by zero-flux-boundary method (\citealt{Duez_Mathis_2010}; 
\citealt{Yoshida_Kiuchi_Shibata_2012})
have succeeded in obtaining the magnetized 
equilibria with strong toroidal 
magnetic fields by choosing very special boundary 
conditions for the poloidal magnetic fields. 
The boundary condition adopted by 
\cite{Tomimura_Eriguchi_2005} 
and \cite{Ciolfi_et_al_2009} in which they 
failed to obtain configurations with 
strong toroidal magnetic fields is that 
the poloidal magnetic field lines should continue 
smoothly through the stellar surfaces into
the vacuum region which is considered to be
outside of the stars. On the other hand, the 
boundary condition employed by
\cite{Glampedakis_Andersson_Lander_2012} is 
different. The poloidal magnetic field lines 
need not continue smoothly at the 
stellar surfaces, because in some of their models 
it has been allowed for the surface currents
to exist. By specifying such a boundary condition, 
they have succeeded in finding that magnetized
configurations whose total energies of the 
toroidal magnetic fields become much stronger as 
the surface currents are increased.

\cite{Duez_Mathis_2010} and 
\cite{Yoshida_Kiuchi_Shibata_2012} also 
obtained the stationary configurations
with strong toroidal magnetic fields, but 
the boundary condition which they adopted
is of different kind from those mentioned above.
Their assumptions are essentially the same as 
those in the previous works (\citealt{Prendergast_1956};
\citealt{Woltjer_1959a, Woltjer_1959b, Woltjer_1960}; 
\citealt{Ioka_Sasaki_2004}).
They imposed the boundary condition that
the magnetic flux on the stellar surfaces
should vanish, so all of poloidal field lines 
are closed and confined inside the stars and
no poloidal magnetic fields penetrate to the 
vacuum region outside of stars. In their solutions, 
the region where the toroidal magnetic fields exist  
inside the star is much larger than that of any other models.
However, they did not explain the reason why 
the magnetized stars can sustain such configurations
with large toroidal magnetic energies
under their special boundary condition.

In this paper we will deal with magnetized
configurations with large amount of the magnetic
energies in the toroidal fields and present 
the reason why the magnetized stars 
can sustain strong toroidal magnetic 
fields within the stars. As will be shown, we have 
found that the total currents of the magnetized 
stars are important keys to understand 
this problem systematically and the values
of the total currents seem to be deeply related 
to the boundary condition of the magnetic fields. 

It should be noted that for the stationary 
configurations the magnetic fields 
are governed by the Grad-Shafranov equation which 
is of the elliptic type partial differential 
equation for the magnetic flux function. 
Therefore, the solutions of the GS equation are
necessarily strongly depending on the boundary 
condition(s).

In this paper, we use both the numerically
exact non-force-free method (\citealt{Tomimura_Eriguchi_2005})
and the method in which  boundary conditions 
are applied at finite locations from the stellar 
centre (e.g., \citealt{Prendergast_1956}; 
\citealt{Ioka_Sasaki_2004}; \citealt{Duez_Mathis_2010}). 
We will show configurations with negative 
surface currents or with the regions where 
the current become negative can sustain the strong 
toroidal magnetic fields inside the star.
Here the term 'negative' means that the 
currents flow in the opposite direction to
the flow direction of the bulk of the 
interior currents.

\section{Formulation}

Our formulation of the problem \ and the numerical 
methods are essentially the same as that of 
\cite{Tomimura_Eriguchi_2005}, i.e. the 
numerically exact method, and that of \cite{Duez_Mathis_2010}, i.e. the zero-flux-boundary method.

\subsection{Grad-Shafranov equation}

We calculate self-gravitating, axisymmetric,
stationary magnetized stars in order 
to obtain magnetized equilibria with strong 
toroidal fields in the Newtonian gravity. We assume 
that the system is in a stationary and axisymmetry 
state. For rotating stars, the rotational axis and 
the magnetic axis coincide and the rotation is 
assumed to be rigid. The star has no meridional flows.
The conductivity of the stellar matter is infinite, 
i.e. the ideal MHD approximation is employed. 
There is no magnetosphere around the star.
In other words, no electric current exists in the 
vacuum  region. 
Therefore, the toroidal magnetic field is confined  
within the star and
the only poloidal component can penetrate the surface 
and extend to the outside of the star.
We use the polytropic equation of state as below:
\begin{eqnarray}
   p = K \rho^{1 + 1/N}.
\end{eqnarray}
Here $p$, $\rho$, $K$ and $N$ are the pressure, 
the mass density, the polytropic constant and 
the polytropic index, respectively.
We fix $N=1$ for simplicity when we compute 
stationary configurations by using the  
numerically exact method.
This choice of $N$ is the same as  that 
adopted in the  previous works 
(e.g. \citealt{Lander_Jones_2009}).

We introduce the magnetic flux function $\Psi$ 
in order to obtain magnetized equilibria efficiently. 
The magnetic flux function is defined as below:
\begin{eqnarray}
   B_r \equiv \frac{1}{r^2 \sin \theta }\P{\Psi}{\theta} , \hspace{10pt}
   B_\theta \equiv - \frac{1}{r \sin \theta }\P{\Psi}{r} \ ,
\end{eqnarray}
where $B_r$ and $B_\theta$ are the components of
the magnetic field in the $r$-direction
and the $\theta$-direction, respectively. 
Using the flux function $\Psi$, 
we derive the Grad-Shafranov equation from Maxwell 
equations as follows:
\begin{eqnarray}
 \Delta^* \Psi \equiv \PP{\Psi}{r} + \frac{\sin \theta}{r^2} \P{}{\theta}
\left(\frac{1}{\sin \theta} \P{\Psi}{\theta}  \right)  = - 4\pi r \sin \theta \frac{j_\varphi}{c},
\label{Eq:GS}
\end{eqnarray}
where $j_{\varphi}$ and $c$ are the 
$\varphi$-component of the current and 
the speed of light, respectively.

The magnetic flux function $\Psi$\ can be expressed as:
\begin{eqnarray}
    \Psi = r \sin \theta A_{\varphi} \ , 
     \label{Eq:Psi_Aphi}
\end{eqnarray}
where $A_{\varphi}$ is the $\varphi$-component of the
vector potential which satisfies 
$\nabla \times \Vec{A} \equiv \Vec{B}$.
The form of Grad-Shafranov equation can
be rewritten as: 
\begin{eqnarray}
   \Delta (A_\varphi \sin\varphi) 
    = -\frac{4 \pi}{c} j_\varphi \sin\varphi,
   \label{Eq:Poisson_A}
\end{eqnarray}
where, $\Delta$ denotes the ordinary Laplacian 
operator. 
This equation seems strange because the system treated in this
paper is axisymmetric and $\varphi$-dependency would not
appear. The reason why we introduce the term $\sin \varphi$
is that by introducing the $\sin \varphi$ term we can get
an equation which contains the three dimensional 
Laplacian operator as shown above. 
The tree dimensional Laplacian operator which emerges
by this seemingly 'strange' device allows us to
transform the equation into an integral form as shown
below.
By taking  the boundary condition for
the vector potential into account and
using Green's function which satisfies
the boundary condition, we derive the integral form 
of the GS equation as follows:
\begin{eqnarray} 
   A_\varphi(\Vec{r}) \sin \varphi
   = \frac{1}{c} \int \frac{j_\varphi(\Vec{r}')}
   {|\Vec{r} - \Vec{r}'|} \sin \varphi' \, d^3 \Vec{r}'
   + A_{\varphi, h} (r, \theta) \sin \varphi,
\label{Eq:int_j_phi}
\end{eqnarray}
where $A_{\varphi,h}$ is a homogeneous general solution 
to Eq.(\ref{Eq:Poisson_A}) as follows:
\begin{eqnarray}
 A_{\varphi,h} (r, \theta) = \sum_{n=0}^{\infty} 
\Big[ a_n \frac{r_0^{n+2}}{r^{n+1}} 
+ b_n \frac{r^n}{r_0^{n-1}} \Big] P_n^1 (\cos \theta).
\label{Eq:homogenous}
\end{eqnarray}
Here, $r_0$ is a certain constant which is the
stellar radius for spherical configurations, and 
$a_n$ and $b_n$ are constant coefficients
which are obtained by applying the boundary 
condition. We will be able to obtain 
the stationary distributions of the magnetic 
vector potentials by solving this equation.
Since these equations are of elliptic type partial 
differential equations whose source term 
is $j_\varphi$, the boundary conditions are very 
important and have significant influences
on the global structures of the
vector potentials or the magnetic flux 
functions. We will deal with this 
problem about the boundary condition in 
Sec. \ref{Sec:surface_current}.

\subsection{Toroidal magnetic fields}

Once we have obtained the flux function by 
solving the GS equation, it is easy to 
calculate the poloidal magnetic fields directly.
On the other hand, we can obtain the toroidal 
component of the magnetic field by using the 
conserved quantity along the flux function
which can be expressed by an arbitrary 
function of $\Psi$. This arbitrary function 
appears because of the assumption of  the axisymmetry.
This function is called $\kappa(\Psi)$ in 
\cite{Tomimura_Eriguchi_2005}, $F$ 
in \cite{Duez_Mathis_2010} 
and $T$ in \cite{Glampedakis_Andersson_Lander_2012}.
In this paper we name it $\kappa(\Psi)$ after 
\cite{Tomimura_Eriguchi_2005}.
The toroidal component of the magnetic field is 
obtained from the following relation 
\begin{eqnarray}
 B_\varphi  = \frac{\kappa(\Psi)}{r \sin \theta}.
\end{eqnarray}
This arbitrary function also appears in the
expression of the current density 
as follows:
\begin{eqnarray}
  \frac{\Vec{j}}{c} = \frac{1}{4\pi} \D{\kappa(\Psi)}{\Psi} \Vec{B} + \rho r \sin \theta
   \mu(\Psi)  \Vec{e}_\varphi \ ,
   \label{Eq:current} 
\end{eqnarray}
where, $\mu(\Psi)$ is another arbitrary function 
of $\Psi$. The arbitrary function $\mu(\Psi)$ is 
the same as  
$G$ in \cite{Duez_Mathis_2010} and 
$F$ in \cite{Glampedakis_Andersson_Lander_2012}.
Then, we express the $\varphi$ component of the
current density as below:
\begin{eqnarray}
 \frac{j_\varphi}{c} = \frac{1}{4\pi} \D{\kappa(\Psi)}{\Psi} \frac{\kappa(\Psi)}{r \sin \theta} + \rho r \sin \theta
  \mu(\Psi) \ .
  \label{Eq:current_phi}
\end{eqnarray}
The first term of Eq.(\ref{Eq:current_phi}) is 
the force-free current density part and 
the second term is the non-force-free current part
because of $\nabla \times \Vec{B} = 4\pi \Vec{j} / c$.
If $\mu = 0$ in a certain region, the magnetic 
field there is force-free, because $\Vec{j} \propto  
\alpha \Vec{B}$ in that region.
We will call the first term as the $\kappa$ term, 
and the second term as the $\mu$ term
of the current density in this paper. 
In a naive treatment, it seems to be enough 
to make the contribution from the  $\kappa$ 
term larger in order to make the toroidal magnetic 
fields larger. However, it has been 
very  difficult to make the influence from 
the $\kappa$ term strong. The reason for that
is as follows. The distribution of $\Psi$ 
is obtained by solving the GS equation, but 
the source term of the GS equation 
contains the $\kappa$ term which is an 
arbitrary function of $\Psi$ and is
confined to a restricted region in the
interior of the star because we impose
the magnetic flux function to be 
{\it smoothly} connected on the stellar 
surface. Therefore, if we change the functional
form and values of parameters which appear
in the functional form of  $\kappa$
as well as the the region where $\kappa$
does not vanish, 
the distribution of the 
flux function also changes according to the 
changes of $\kappa$. 
We will discuss this difficulty in 
Sec.\ref{Subsec:no_surface_current_model}.

%
 \section{Surface currents}
\label{Sec:surface_current}

In this section, we deal with the relation 
between the surface current and the magnetic 
field, which is deeply influenced by the 
boundary conditions. The surface current
can be defined either by the discontinuity
of the derivative of the magnetic flux function, 
or by the homogeneous term in the integral 
representation for the vector potential.
The both definitions for the surface current 
give exactly the same values as we will see
later.

 \begin{figure*}
 \includegraphics[scale=0.83]{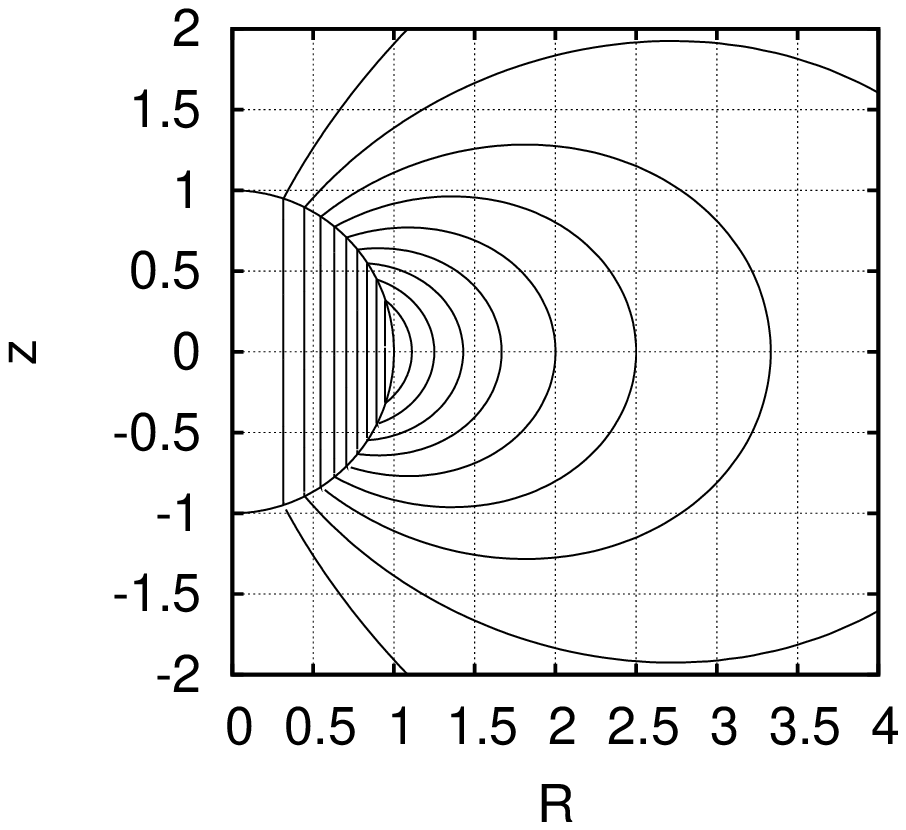}
 \includegraphics[scale=0.83]{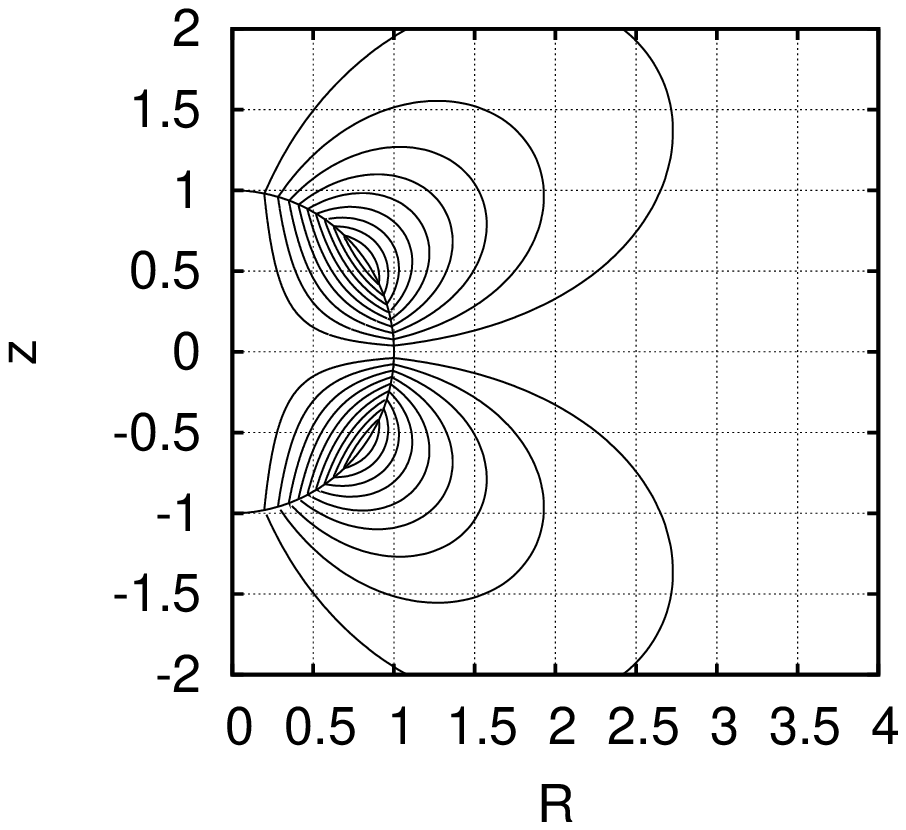}
 \caption{Poloidal magnetic field lines with a
purely dipole (n=1) surface current (left) and
a purely quadrupole $(n=2)$ surface current 
(right).}
 \label{Fig:dipole_quadrupole}
 \end{figure*}

\subsection{Relation between the surface current 
and the discontinuity of the magnetic field}
\label{Sec:j_sur_diff}

At first, we will show a relation between the 
surface current and the 
discontinuity of the magnetic field as follows.
If the Amp\'{e}re's equation is applied to an area
$S$ bounded by a boundary $\partial S$, we can write
it by an integral form as follows:
\begin{eqnarray}
 \oint_{\partial S} \Vec{B} \cdot d\Vec{\ell} 
= \frac{4\pi}{c} \int_{S} \Vec{j} \cdot d \Vec{S},
\end{eqnarray}
where $d\Vec{\ell}$ and $d\Vec{S}$ are a line element 
and a surface element, respectively.
We apply this equation to an infinitely small 
area in the meridional plane of the star bounded 
by four lines as follows:
\begin{eqnarray}
r & = & r_s(\theta_1) - \frac{\Delta r}{2} \ , \\
r & = & r_s(\theta_1) + \frac{\Delta r}{2} \ , \\
\theta & = & \theta_1 \ , \\
\theta & = & \theta_1 + \Delta \theta \ ,
\end{eqnarray}
where $\theta_1$, $\Delta r$ and $\Delta \theta$
are a constant, infinitesimal widths in the
$r$-direction and in the $\theta$-directions, 
respectively. For this infinitesimal area,
we obtain
%
{\small
\begin{eqnarray}
&& \Big[B_r^{ex}(r_s, \theta) 
     - B_r^{in}(r_s,\theta)\Big] \Delta r 
+ \Big[B_\theta^{ex}(r_s, \theta) 
     - B_\theta^{in}(r_s, \theta)  \Big] 
     r_s \Delta \theta
\nonumber \\
  &=& \frac{4\pi}{c} 
    \int_{r_s -\Delta r/2}^{r_s + \Delta r/2}
       j_{\varphi}(r, \theta) r \, dr 
         \Delta \theta,
\label{app:dis_j_sur}
\end{eqnarray}
}
%
where $B^{ex}$ and $B^{in}$ are the exterior and 
interior values of  the magnetic fields. 
Here, if the current density is a surface current on the 
stellar surface $(r = r_s(\theta))$ defined by
$
 j_{\varphi}(r,\theta) = j_{sur}(\theta) \delta(r-r_s(\theta))
$,
we can integrate the equation as below

\begin{eqnarray}
   && \int_{r_s -\Delta r/2}^{r_s + \Delta r/2}
       j_{\varphi}(r, \theta) r \, dr \nonumber \\
&=&    \int_{r_s -\Delta r/2}^{r_s + \Delta r/2}
       j_{sur}(\theta) \delta(r-r_s) r \, dr 
=  r_s j_{sur} (\theta).
\end{eqnarray}
Since the $r$ dependence  of the magnetic fields are 
continuous on the stellar surface ($r = r_s(\theta)$), 
we obtain a relation as follows:
\begin{eqnarray}
 B_\theta^{ex}(r_s, \theta) - B_\theta^{in}(r_s, \theta) = \frac{4\pi}{c} j_{sur}(\theta).
\end{eqnarray}
The surface current is expressed by the discontinuity 
of the $\theta$-component of the 
magnetic field. For more general situations, 
we can obtain the following equation for the
surface current using the parallel component 
$B_{||}$ to the stellar surface,
\begin{eqnarray}
 B_{||}^{ex}(r_s, \theta) - B_{||}^{in}(r_s, \theta) = \frac{4\pi}{c} j_{sur}(\theta).
\label{Eq:b4}
\end{eqnarray}
If the surface current exits, 
the parallel component of the magnetic field 
must be discontinuous. 
We emphasize that the value of the discontinuity 
of the magnetic field between just inside
and just outside of the stellar surface equals 
the surface current density.

\cite{Glampedakis_Andersson_Lander_2012} expressed 
the surface current density in a different way as
follows. They defined the surface current by 
imposing the discontinuity of poloidal magnetic 
fields at the stellar surface (see Eq.67 
in \citealt{Glampedakis_Andersson_Lander_2012}). 
Their discontinuous boundary condition is
just an {\it assumption} without a firm 
foundation as
follows:
\begin{eqnarray}
   b^{in}_\theta  = \xi b^{ex}_\theta  \Leftrightarrow b_\theta^{ex} - b_\theta^{in} = (1-\xi) b_\theta^{ex} = j_{sur}.
\label{Eq:b5}
\end{eqnarray}
Here $b$ and $\xi$ indicate the magnetic field 
and a discontinuity parameter, respectively,
in their paper. 
Since their model is a purely dipole configuration,  
the exterior solution of $b_\theta$ is 
$b_\theta^{ex}(x) = \frac{1}{2x} 
d\hat{\alpha}/dx \sin \theta$ (Eq.61 in 
\citealt{Glampedakis_Andersson_Lander_2012}).
Here $\alpha$ and $x$ indicate the $r$-component 
of the flux function and the dimensionless radius 
normalized by the stellar radius.
Then we can calculate the distribution of their surface 
current density as follows,
\begin{eqnarray}
 j_{sur} = \left[ (1 - \xi )\frac{1}{2x}\D{\alpha}{x} \sin \theta \right]_{(x=1)} = - j_0 \sin \theta , 
\end{eqnarray}
where 
\[
j_{0} = \left[ (\xi - 1)\dfrac{1}{2x} \P{\hat{\alpha}}{x} \right]_{(x=1)}.
\]
Since they calculated only models with $\xi > 1$
in their paper, the surface current density of 
their models flows in the opposite direction to
the interior bulk toroidal current density inside the star. 
In other words their surface current 
is negatively flowing with respect to the
bulk interior currents. According to 
\cite{Glampedakis_Andersson_Lander_2012}, if the 
value for the discontinuity for the poloidal
magnetic field is increased, the energy of the  
toroidal magnetic fields becomes larger 
(see Fig. 5 in 
\citealt{Glampedakis_Andersson_Lander_2012}).
Therefore we conclude that the negative surface 
current sustain strong toroidal
magnetic fields comparing with those in 
Tomimura \& Eriguchi models without 
surface currents.

\subsection{Surface currents in the integral 
representation}
\label{Sec:j_sur_int}

Using the integral representation for the
vector potential, we can see the surface current 
from a different point of view. We assume that a 
magnetized star has no currents 
in the stellar interior except for the surface current.
It implies that the source term for the GS equation 
consists  only of the surface current.
We can obtain the magnetic field by  calculating 
Eq. (\ref{Eq:int_j_phi}),
\begin{eqnarray}
 A_\varphi \sin \varphi = \frac{1}{c}\int \frac{j_{sur}(r', \theta')}{|\Vec{r} - \Vec{r}'|} \sin \varphi' \, d^3 \Vec{r}' .
 \label{Eq:c1}
\end{eqnarray}
Since the surface current  exists on the 
stellar surface $r=r_s$,  we can describe the 
surface current density using the Dirac's delta 
function:
\begin{eqnarray}
 \frac{j_{sur}(r', \theta')}{c} 
=  \delta(r' - r_s) j_{sur}(\theta') \ , 
\end{eqnarray}
where $j_{sur} (\theta)$ is the surface current
which flows along the surface.
We can expand and integrate Eq.(\ref{Eq:c1}) 
using the Legendre functions and the 
axisymmetry of the system. We obtain 
the solutions for $A_\varphi$ as follows:
\begin{eqnarray}
A_\varphi (r, \theta) &=&
\sum_{n=1}^{\infty} \dfrac{2\pi}{n(n+1)}P_n^1 (\cos \theta) f_n(r,r_s) r_s^2 \nonumber \\
&&
\int_0^{\pi} \sin \theta' P_n^1 (\cos \theta') j_{sur}(\theta') \, d\theta'  ,
\label{Eq:c3}
\end{eqnarray}
where $f_n$ is a function defined by
\begin{eqnarray}
f_n(r,r_s) = 
\left\{
     \begin{array}{lr}
      r_s^n/r^{n+1} , & (r\geq r_s) \\
      r^n/r_s^{n+1} . & (r\leq r_s)
    \end{array}
  \right .
\end{eqnarray}
Now we calculate the vector potential and 
the magnetic field by giving a 
$\theta $-distribution for the surface current density.
We assume the $\theta$-distribution of 
the surface current can be expressed by the 
expansion using Legendre functions,
\begin{eqnarray}
 j_{sur} (\theta) = \sum_{n=1}^{\infty} 
\frac{2n + 1}{4\pi}\alpha_n P_n^1 (\cos \theta),
\end{eqnarray}
where $\alpha_n$'s are dimensionless 
coefficients related to the $n$th 
associate Legendre function $P_n^1(\cos \theta)$.
Then using the orthogonality among 
the Legendre functions,\
\begin{eqnarray}
 \int_0^{\pi} \sin \theta' P_n^1 (\cos \theta ') P_n^1(\cos \theta ') \, d \theta' = 2\frac{(n+1) n }{2n + 1},
\end{eqnarray}
we can obtain an analytically expressed 
solutions as follows:
\begin{eqnarray}
A_\varphi(r, \theta) =
\sum_{n=1}^{\infty} \alpha_n r_s^2 f_n(r,r_s) P_n^1(\cos \theta),
\label{Eq:Aphi_anal}
\end{eqnarray}
and 
\begin{eqnarray}
\Psi(r, \theta) =
\sum_{n=1}^{\infty} \alpha_n r r_s^2 f_n(r,r_s) P_n^1(\cos \theta) \sin \theta.
\label{Eq:Psi_anal}
\end{eqnarray}
If we set $\alpha_n = a_n + b_n$,
the right hand side of Eq. (\ref{Eq:Aphi_anal}) 
is exactly the same as homogeneous general 
solutions of Eq. (\ref{Eq:homogenous}).   
Therefore, adding the homogeneous term to the
inhomogeneous solution of the GS equation
corresponds exactly to adding the surface 
current at the boundary surface. 

Fig. \ref{Fig:dipole_quadrupole} shows the 
poloidal magnetic field lines for 
configurations with the 
purely dipole ($n=1$) surface current 
(left panel) and with the purely quadrupole 
($n=2$) surface current (right panel).
It should be noted that each model has
no interior currents except for 
the surface currents
as we have described in this section.

 \section{Numerically exact configurations for open magnetic fields
with surface currents}

We will show numerically exact structures of
magnetized stars with open field lines
and with/without surface currents in 
this section. At first, we will display 
configurations which have no surface currents.
Although they are the same as those obtained in 
\cite{Yoshida_Eriguchi_2006} and
\cite{Lander_Jones_2009}, we will check these 
models from a different point of view, i.e. 
in the context of the influence of the surface 
currents. Next, magnetized stars which have
surface currents will be treated and discussed. 

\subsection{Setting of the problem}

As discussed before, we have solved the 
integral equation derived from the GS equation
by considering the presence of the surface 
currents as follows:
\begin{eqnarray} 
   A_\varphi(\Vec{r}) \sin \varphi
   = \frac{1}{c} \int \frac{j_\varphi(\Vec{r}') + j_{sur}(\Vec{r}')}
   {|\Vec{r} - \Vec{r}'|} \sin \varphi' \, d^3 \Vec{r}' \ ,
\label{Eq:int_j_phi2}
\end{eqnarray}
where $j_{sur}$ is the surface current density 
of the magnetized star. We choose the following
two different distributions for the 
surface currents:
{\small
\begin{eqnarray}
 \frac{j_{sur}(r,\theta)}{c} = - j_0 \sin \theta \delta(r - r_s(\theta)),
\ (\mbox{dipole distribution}) \ ,
  \label{Eq:j_sur_dip}
\end{eqnarray}
}
and 
{\small
\begin{eqnarray}
 \frac{j_{sur}(r,\theta)}{c} = - j_0 \sin \theta \cos \theta \delta(r - r_s(\theta)). \  (\mbox{quadrupole distribution}) \ 
  \label{Eq:j_sur_quad}
\end{eqnarray}
}

As for the arbitrary functions appearing in our
formulation, we choose the following forms: 
\begin{eqnarray}
 \mu(\Psi)   &=& \mu_0  \ , \\ 
 \int    \mu(\Psi) \, d \Psi     &=& \mu_0 \Psi \ ,
\label{Eq:mu}
\end{eqnarray}
\begin{eqnarray}
\kappa (\Psi)   = 
\left\{
\begin{array}{ll}
0, &
 \mathrm{for} \ \Psi \leq \Psi_{V\max} \ , \\
 \dfrac{\kappa_0}{k + 1} (\Psi - \Psi_{V\max})^{k + 1},  & 
\mathrm{for} \ \Psi > \Psi_{V\max} \ ,
\label{Eq:kappa}
\end{array}
\right.
\end{eqnarray}
and 
\begin{eqnarray}
 \kappa'(\Psi)  = 
\left\{
\begin{array}{ll}
0, & \mathrm{for} \ \Psi \leq \Psi_{V\max} \ , \\
\kappa_0(\Psi - \Psi_{V\max})^{k}, & 
\mathrm{for} \ \Psi > \Psi_{V\max} \ , \\
\end{array}
\right.
\end{eqnarray}
Here, $\mu_0$, $k$ and $\kappa_0$ are constant 
parameters and $\Psi_{V \max}$ means the maximum 
value of $\Psi$ in the vacuum region. In this paper,
we fix $k = 0.1$. These functional forms and 
the value of $k$ are the same as those 
chosen in other papers
(\citealt{Yoshida_Eriguchi_2006}; \citealt{Lander_Jones_2009}).
In this section, we set the polytropic index  
$N=1$ (e.g. \citealt{Lander_Jones_2009}), and 
$q=0.9$ where $q$ is the ratio of the polar radius 
to the equatorial radius defined by
(see \citealt{Fujisawa_Yoshida_Eriguchi_2012})
\begin{eqnarray}
 q \equiv \frac{r_s(\theta = 0)}{r_s(\theta=\pi/2)}.
\end{eqnarray}
Concerning the angular velocity $\Omega$, we 
choose values of $\hat{\Omega} = 
\mbox{constant}\equiv \hat{\Omega}_0 = 1.0$E-2  
for rigidly rotating configurations and 
$\hat{\Omega}_0 = 0.0$ for
non-rotating models. Here quantities with $\hat{}$ 
represent the corresponding ones transformed
into dimensionless forms as shown in 
Appendix \ref{App:Dimensionless}
(see also \citealt{Fujisawa_Yoshida_Eriguchi_2012}). 

Although the equations of state influence the 
strengths of the toroidal magnetic fields
(see \citealt{Kiuchi_Kotake_2008}), 
we have treated only $N = 1$ polytropes because
the main concern in this paper is how the surface 
current density affects the distributions of
the magnetic fields, in particular to the
toroidal magnetic fields. 
In order to examine the accuracies of solutions, 
we have used the virial relation as
shown in Appendix \ref{App:VC}.

\subsection{Configurations without surface currents}
\label{Subsec:no_surface_current_model}

\begin{figure*}
  \begin{center}
   \includegraphics[scale=0.64]{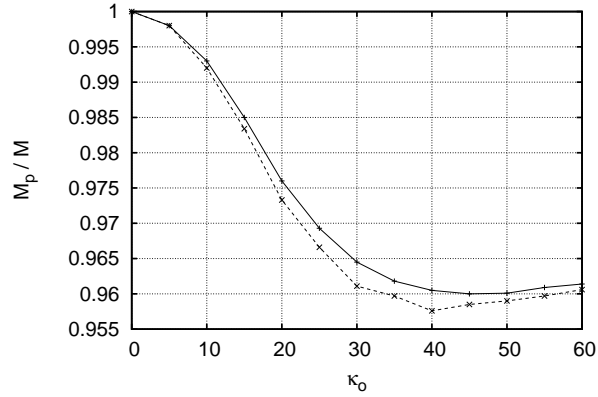} 
  \end{center}
 \caption{The ratio of the poloidal magnetic energy to 
 the total magnetic energy against the
 parameter $\hat{\kappa}_0$.
 The solid line denotes a rigidly rotating sequence 
 with $\hat{\Omega}_0 = 1.0$E-2.0
 and the dashed line denotes a non-rotating 
 sequence, i.e. a sequence with 
 $\hat{\Omega}_0 = 0.0$. Minimum values seem to
 be attained  at $\hat{\kappa}_0 \sim 40$ (for the
 non-rotating sequence) and $\hat{\kappa}_0 
 \sim 45$ (for the rigidly rotating sequence).
}
\label{Fig:Mp_M0}
\end{figure*}

\begin{figure*}
   \includegraphics[scale=0.68]{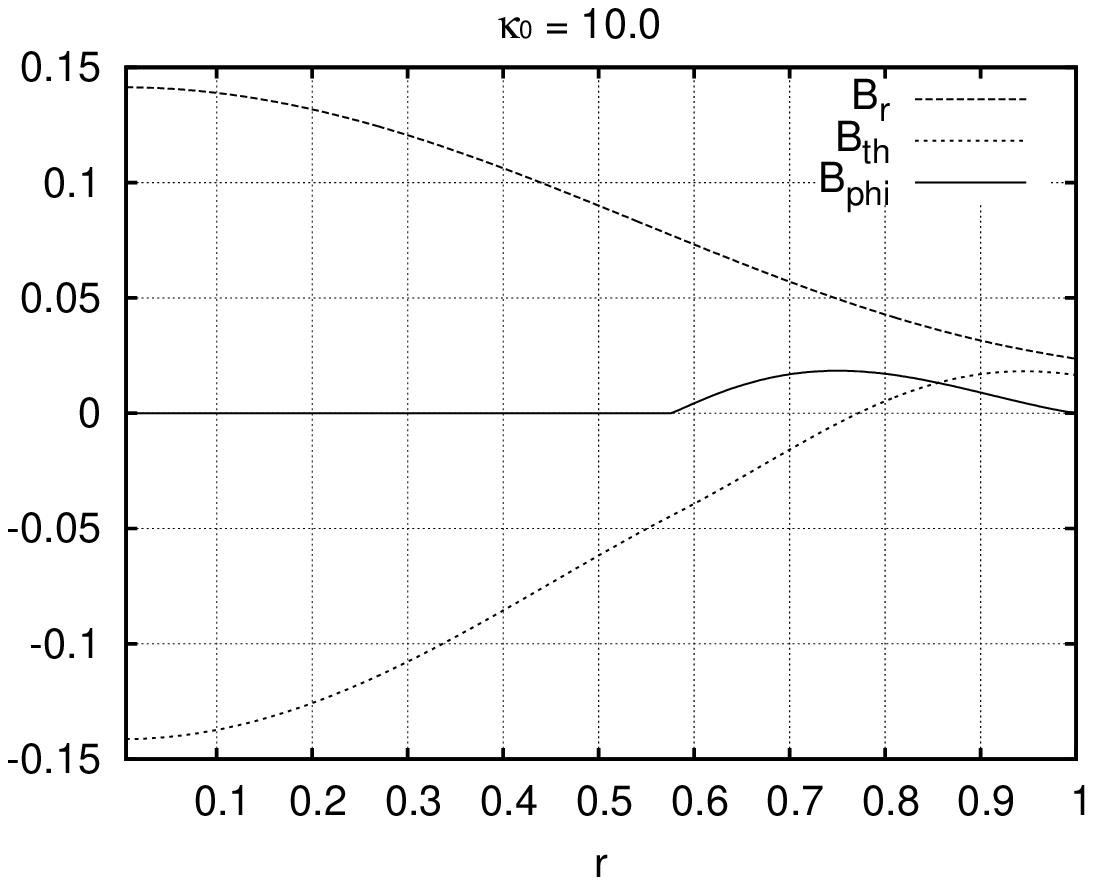}
   \includegraphics[scale=0.68]{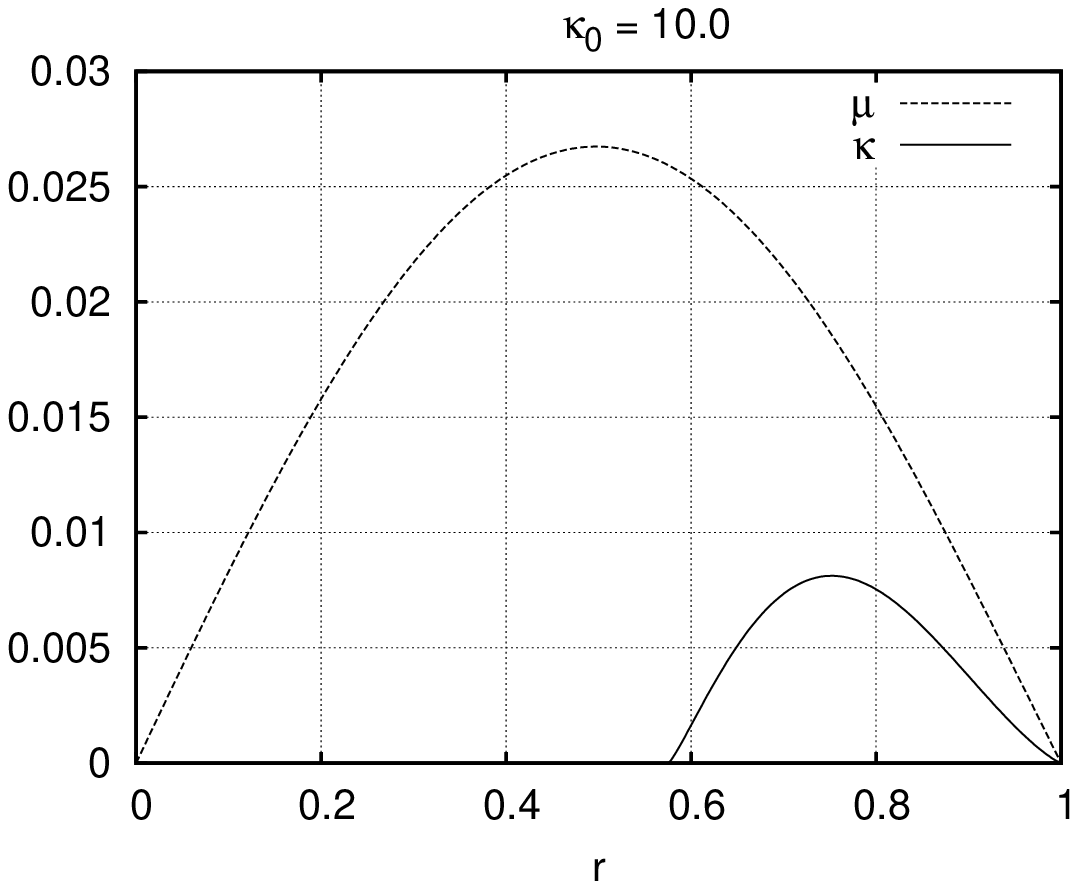}

   \includegraphics[scale=0.68]{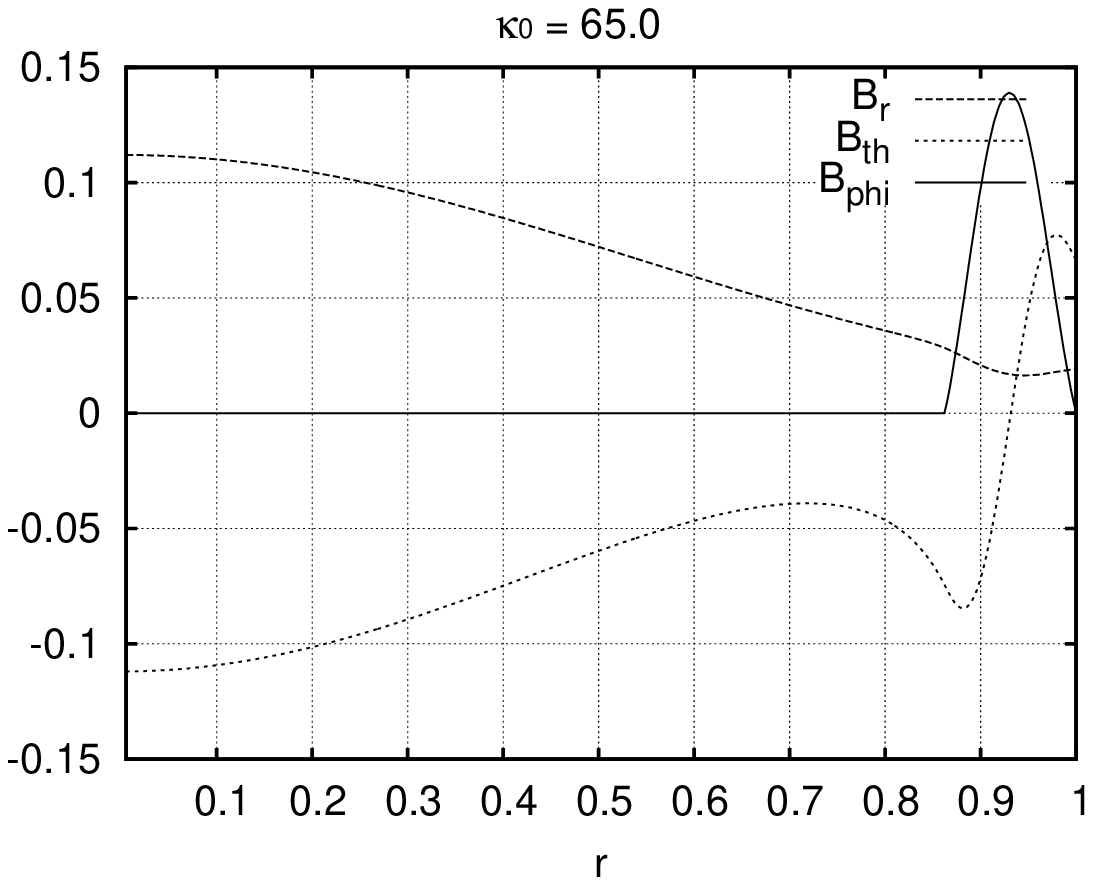}
   \includegraphics[scale=0.68]{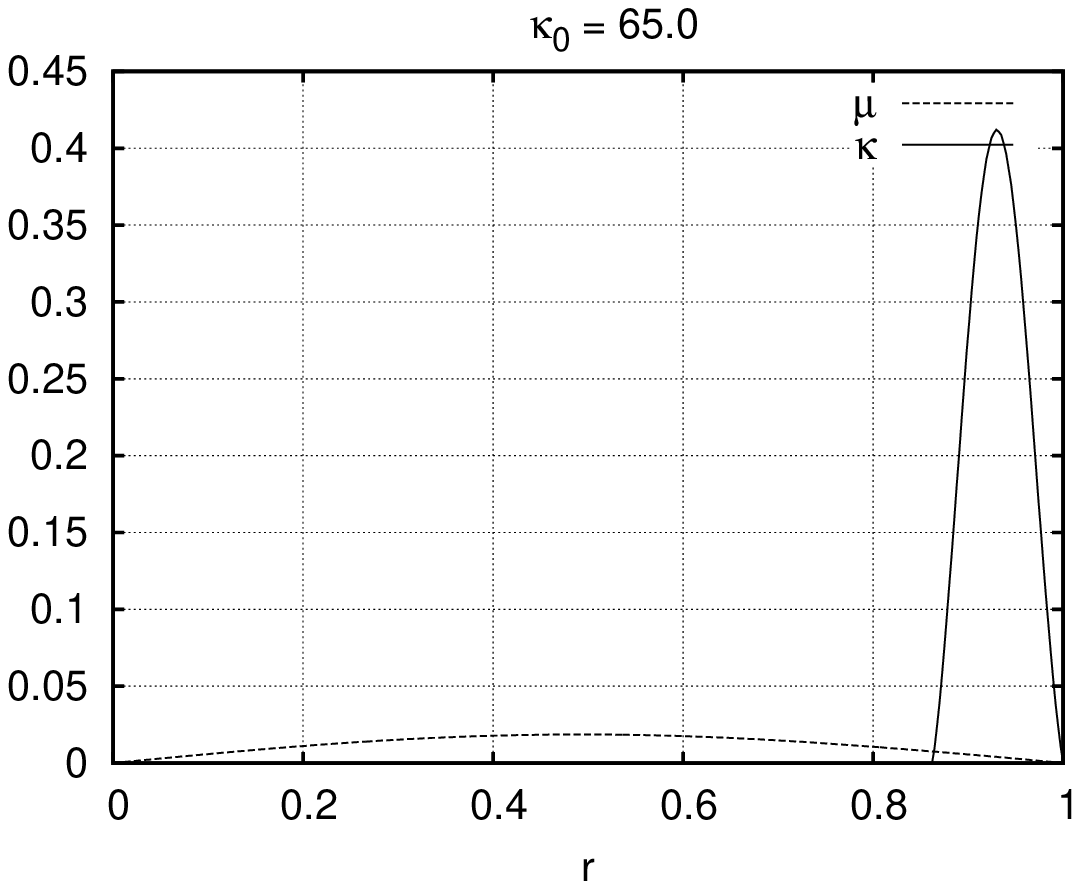}
 \caption{Left panels: Distributions of the 
 components of the magnetic field 
 [$\hat{B}_r (\theta =0), \hat{B}_\theta (\theta = \pi / 2), 
 \hat{B}_\varphi (\theta = \pi / 2)$] are shown. 
Right panels : Distributions 
 of the two source terms on the equator, which 
 appear in the current density formula, are 
 shown. Chosen values of $\hat{\kappa}$ are 
 $\hat{\kappa}_0 = 10$ (top) 
 and $\hat{\kappa}_0 = 65$ (bottom).
\label{Fig:H_r_j_r_jsur0}
}
\end{figure*}

Since the value of 
$\hat{\kappa}_0$ affects the local 
behavior of the toroidal magnetic field
distributions, in particular on its maximum 
value (see \citealt{Lander_Jones_2009}), 
we have solved the magnetized configurations
by changing the value of $\hat{\kappa}_0$
for two values of $\hat{\Omega}_0$.

Obtained results are plotted in 
Fig.\ref{Fig:Mp_M0} which shows the ratio of 
${\cal M}_p / {\cal M}$ against $\hat{\kappa}_0$. 
As seen from this figure, there is a minimum value 
of the ratio.  It implies that the toroidal 
magnetic field energy increases to its maximum 
value at $\hat{\kappa_0} \sim 40 $ 
for non-rotating configurations and at 
$\hat{\kappa}_0 \sim 45 $ for rigidly rotating
models. Since the term related to the rotation 
does not appear in the current density 
formula (see Eq.\ref{Eq:current}),
the rotation affects the toroidal magnetic 
field distributions only slightly. Therefore, 
we will display and discuss only configurations
with rotation in the following part of this 
section.

In many investigations which have been done
by applying numerically exact methods 
or by structure separated GS solving method 
zero-flux-boundary methods, almost similar
results as ours shown in Fig. \ref{Fig:Mp_M0} 
have been obtained (see Table 2 
in \citealt{Lander_Jones_2009}, Fig. 12 
in \citealt{Ciolfi_et_al_2009}, Fig. 4 
in \citealt{Glampedakis_Andersson_Lander_2012}).
Therefore,  this behavior of the toroidal
magnetic field against the value of 
$\hat{\kappa_0}$ is likely to be a general 
feature of stationary magnetized stars
which have open magnetic fields.

In order to consider the reason of the 
presence of these minimum values, in 
Fig.\ref{Fig:H_r_j_r_jsur0} shown are 
the distributions of the magnetic field 
components (left panels) and those of
the two components of the current density 
formula (right panels). Different curves in 
the left panels mean 
$\hat{B}_r(\theta=0)$ (dotted line), 
$\hat{B}_\theta(\theta=\pi/2) $(dashed line) 
and $\hat{B}_\varphi (\theta=\pi/2) $(solid 
line) distributions for the region 
$\hat{r} [0:1.0]$. In the right panels, 
the solid line denotes the force-free 
$\kappa$ term and the 
dashed line denotes the non-force-free $\mu$ 
term in the current density formula.

As seen from these panels, by increasing the value 
of $\hat{\kappa}_0$, which corresponds to 
increasing the maximum strength of the toroidal 
magnetic field, from top panels to bottom panels, 
we can find that the width of the 
toroidal magnetic field region becomes narrower. 
The values of the toroidal magnetic field 
energy seem to depend on the distributions 
of the toroidal magnetic 
fields and of the current densities, in
particular, on the maximum value and the 
width of the toroidal magnetic field distribution.
Although the maximum value of the $\kappa$ term of
the  current density increases with 
$\hat{\kappa}_0$, its width becomes 
narrower as $\hat{\kappa}_0$ is increasing. 
It implies that the slope of the distribution of 
the $\kappa$ term becomes steeper for
the large value of the current density. 
This tendency is also seen in the distribution
of $B_\varphi$. By contrast with this, 
the distributions of the $\mu$ term are 
almost the same because we fix $q$ and 
$\hat{\Omega}_0$ which are related to the 
characteristic nature of the non-force-free 
magnetic fields.

From Eq. (\ref{Eq:current}) and our numerical 
results, we can find that in order to sustain 
strong toroidal magnetic fields 
(appearing in the right hand side 
of the equation), the strong toroidal current 
density 
(appearing in the left hand side of the equation) is
required. It should be noted that the strength
of the toroidal magnetic field seems to be
related deeply to the strength of the total
current density. This can be seen from the 
following argument. We introduce several
definitions about integrated currents
as follows:
\begin{eqnarray}
 \hat{J}_{tot}^{(+)} & \equiv & 
   \int_{S_{mer}} \hat{j}_\varphi^{(+)} dS  \ , \\
 \hat{J}_{tot}^{(-)} & \equiv &
   \int_{S_{mer}} \hat{j}_\varphi^{(-)} dS  \ , \\
 \hat{J}_{tot}^{\kappa} & \equiv &
   \int_{S_{mer}} \hat{j}_{\varphi}^{\kappa} dS   \ , \\
 \hat{J}_{tot}^{\mu} & \equiv &
   \int_{S_{mer}} \hat{j}_{\varphi}^{\mu} dS   \ , \\
 \hat{J}_{sur} & \equiv &
   \int_{S_{mer}} \hat{j}_{sur} dS = - \hat{j}_0 \int_0^\pi   r_s(\theta) \sin \theta d \theta   \ , \\
 \hat{J}_{tot} & \equiv &
     \hat{J}_{tot}^{(-)}   +  \hat{J}_{tot}^{(+)} 
       + \hat{J}_{sur} = \hat{J}_{tot}^{\kappa} + \hat{J}_{tot}^{\mu} + \hat{J}_{sur} \ , 
\end{eqnarray}
where {$S_{mer}$} denotes the meridional plane 
which is perpendicular to the
$\varphi$-coordinate and $dS$ is an area element
in the meridional plane.
Here, $\hat{j}_{\varphi}^{(+)}$,
$\hat{j}_{\varphi}^{(-)}$,
$\hat{j}_\varphi^{\kappa}$ and $\hat{J}_{\varphi}^{\mu}$
the $\varphi$-component of the positively flowing 
interior current density,  the $\varphi$-component 
of the negatively flowing interior current density,  
the $\kappa$ term of the current density and the $\mu$ term.
Furthermore, $\hat{J}_{tot}^{(+)}$, $\hat{J}_{tot}^{(-)}$, 
$\hat{J}_{tot}^{\kappa}$, $\hat{J}_{tot}^{\mu}$, 
$\hat{J}_{sur}$ and $\hat{J}_{tot}$  are
the total {\it positive} bulk interior current,
the total {\it negative} bulk interior current,
the total $\kappa$ term bulk interior current,
the total $\mu$ term bulk interior current,
the total surface current and
the total (bulk + surface) current in the meridional
plane, respectively.
As we shall see, these quantities will play key 
roles to understand the problem.

\begin{figure*}
 \begin{center}
 \includegraphics[scale=0.625]{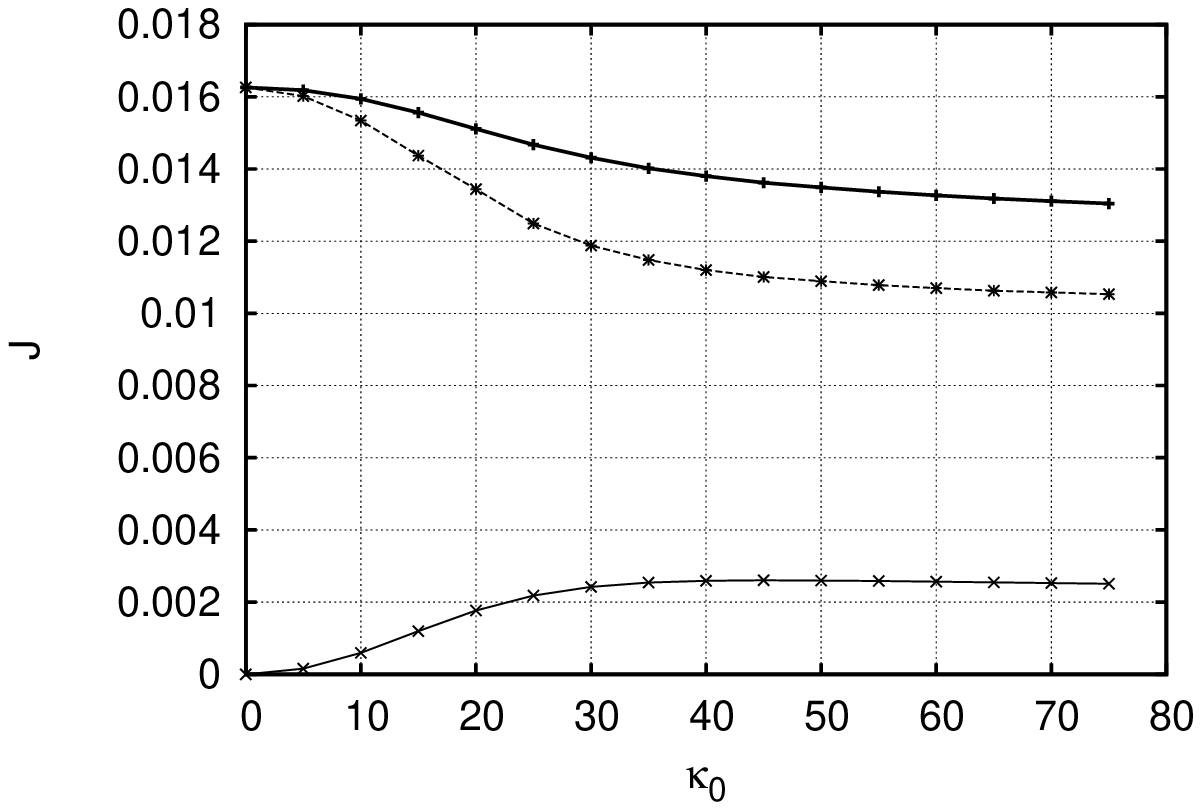}
 \includegraphics[scale=0.625]{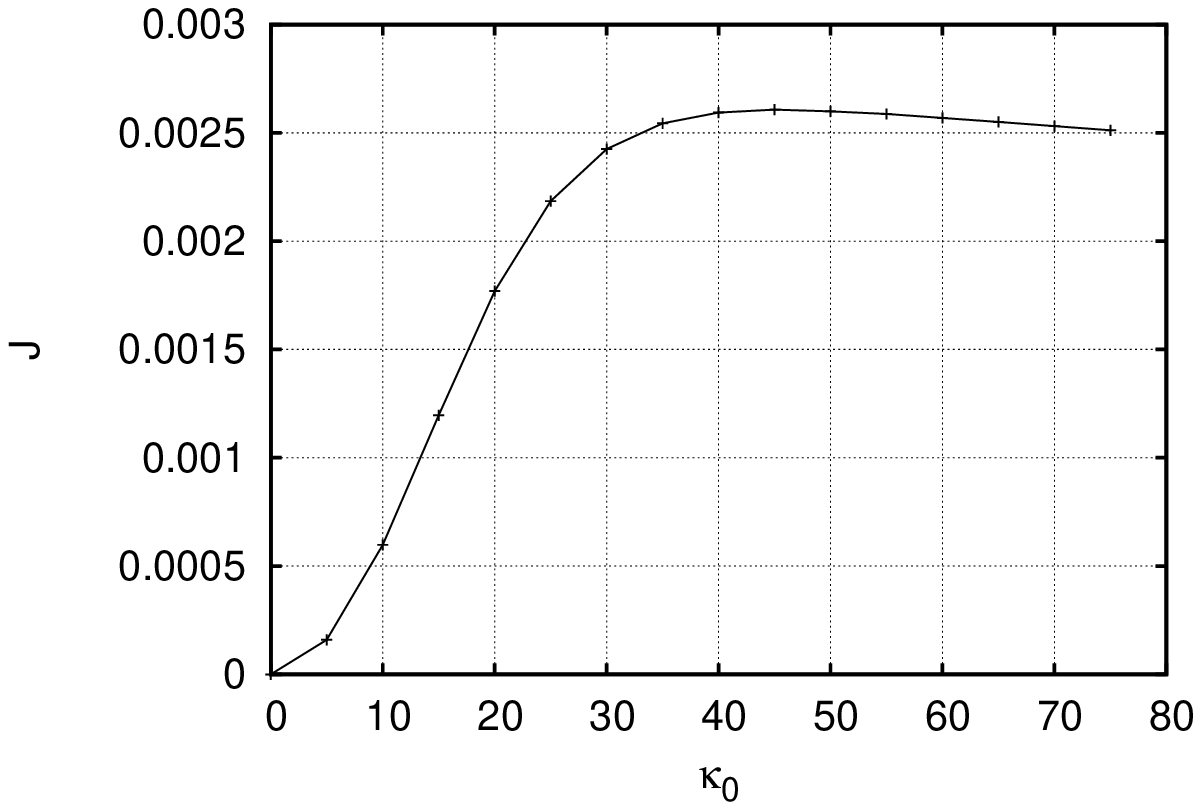}
 \end{center}
\caption{Left panels: The total current $\hat{J}_{tot}$ (thick solid line),
the total $\mu$ term interior current $\hat{J}_{tot}^{\mu}$ (dashed line) 
and the total $\kappa$ term interior current $\hat{J}_{tot}^{\kappa}$ (thin dashed line)
are plotted against $\hat{\kappa}_0$ for configurations
without surface currents. 
Right panels: Only the $\hat{J}_{tot}^{\kappa}$ is plotted against $\hat{\kappa}_0$.
The magnitude of the  $\hat{J}_{tot}^{\kappa}$ has 
an upper bound near at $\hat{\kappa}_0 \sim 45$ .}
\label{Fig:j_phi_j_tot}
\end{figure*}

In Fig.\ref{Fig:j_phi_j_tot}, the total
current, the total $\mu$ current and the total 
$\kappa$ current of the star is plotted against 
$\hat{\kappa}_0$.  We find from Fig. \ref{Fig:j_phi_j_tot} that
the total current does not increase as $\kappa_0$ increases
and the total $\kappa$ current increases to its maximum value 
near at $\hat{\kappa}_0 \sim 45$.  We will denote 
the maximum value of the total current as $J_{tot}^{(max)}$
in this paper. This $\hat{\kappa}_0$ value is the same as that 
for the minimum value of ${\cal M}_p / {\cal M}$.
Therefore it is important to note that there is an upper 
bound of the total current for configurations
if we consider a stationary sequence with different 
values of $\hat{\kappa}_0$.

This upper bound comes from our boundary 
condition for the current density.
Since we have imposed that the outside of the 
star is vacuum, the current density needs to 
vanish in that vacuum region. As we have seen, 
the magnetized stars need large and strong 
toroidal currents in order to sustain strong 
toroidal magnetic fields.  
However, the boundary condition sets limit to
the total current of the star 
as seen from Fig.\ref{Fig:j_phi_j_tot}.
As a result, the region where the current density
attains  a rather large value becomes smaller
and the slope of the distribution of the current
density becomes steeper in order to sustain 
the stronger toroidal magnetic field in the 
narrower region.

Moreover, larger values of 
$\hat{\kappa}_0$
cause the maximum value of the magnetic
flux function in the {\it vacuum} region
larger, in general. As far as our boundary
condition for the magnetic flux function 
to be smooth at the stellar surface is
employed, the support of the $\kappa$
function becomes smaller and smaller as
the value of $\hat{\kappa}_0$ is increased.
In other words, increasing the value
of $\hat{\kappa}_0$ might, in ordinary 
situations, result in increasing the
interior currents but at the same time
decreasing the support region of the
function $\kappa$ because the
maximum value of the magnetic flux 
function in the vacuum region also becomes
larger as explained before. 

This is the reason  why in the present
investigation as well as in other works
thus far done nobody could obtain solutions which 
exceed this upper bound. To overcome this 
limitation about the size of the
confined region of the large toroidal
magnetic field, the magnetized stars needs 
other kinds of distributions for the toroidal 
current densities.

From these consideration, we need to devise
some means to fulfill the following seemingly
contradicted requirements at the same time.

(1) $\varphi$-currents must be increased.
     In ordinary situations, this would
     results in reducing the support region
     of the function $\kappa$ because
     of the increase of the maximum value
     of the magnetic flux function in the
     vacuum.
(2) The support region for the function
    must be widened. In ordinary situations,
    the support region of the function
    $\kappa$ is wider for the smaller
    values of $\kappa_0$.

These two seemingly contradictory requirements
could be realized by introducing
{\it negatively flowing} currents
near/on the surface because the negatively 
flowing currents allow the positively
flowing interior currents to become
larger and at the same time
negatively flowing currents near/on the
surface could reduce the value of the
magnetic flux function in the
vacuum region and result in the smaller 
value for $\Psi_{Vmax}$.

\subsection{Configurations with surface currents  
-- Dipole currents}

\begin{figure*}
 \includegraphics[scale=0.64]{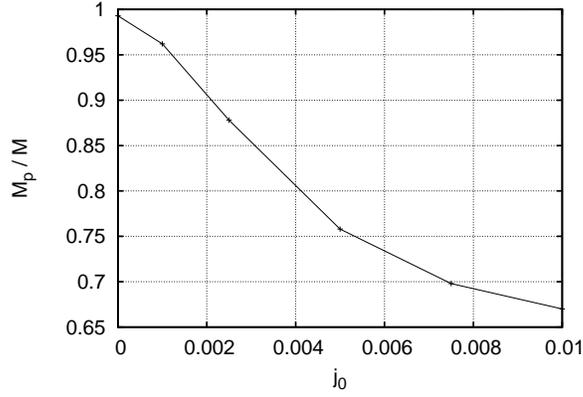}
 \caption{The ratio of the poloidal magnetic energy to
the total magnetic energy of the models with surface current 
against the parameter $\hat{j}_0$.
}
\label{Fig:j0}
\end{figure*}

\begin{figure*}
 \includegraphics[scale=0.62]{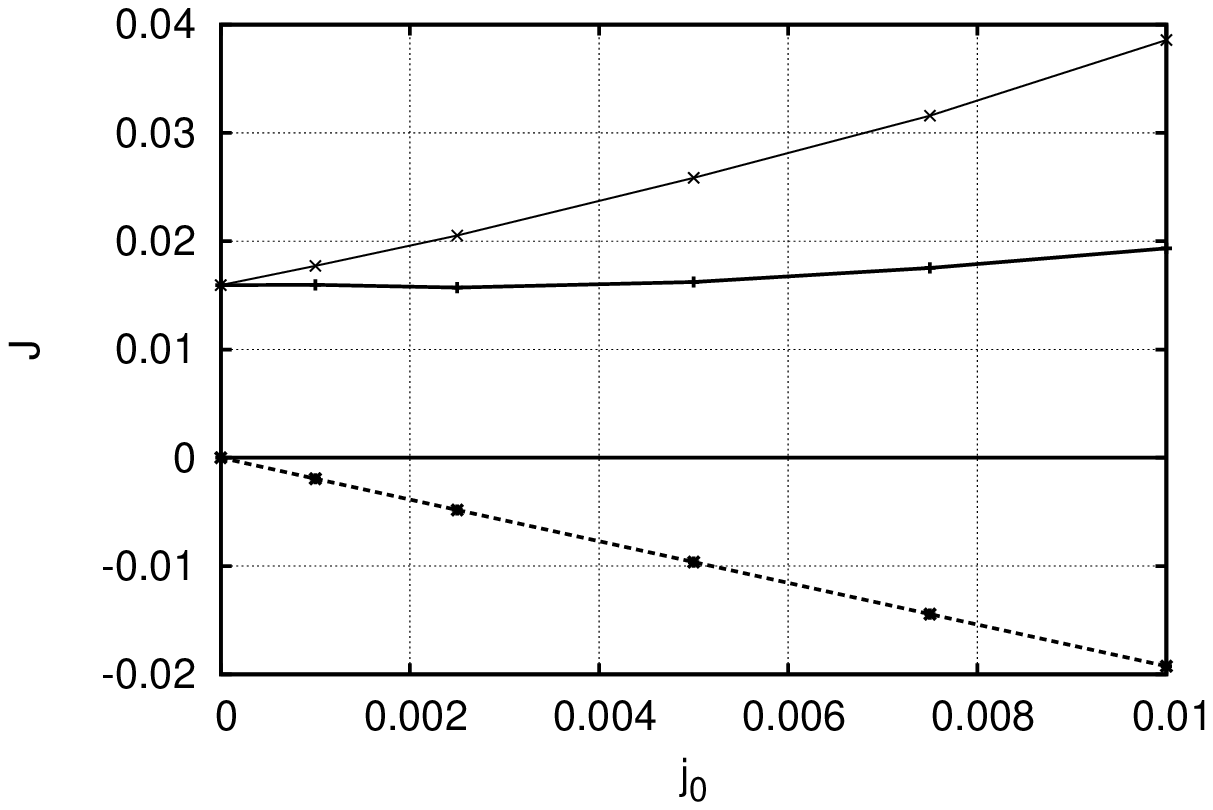}
 \includegraphics[scale=0.62]{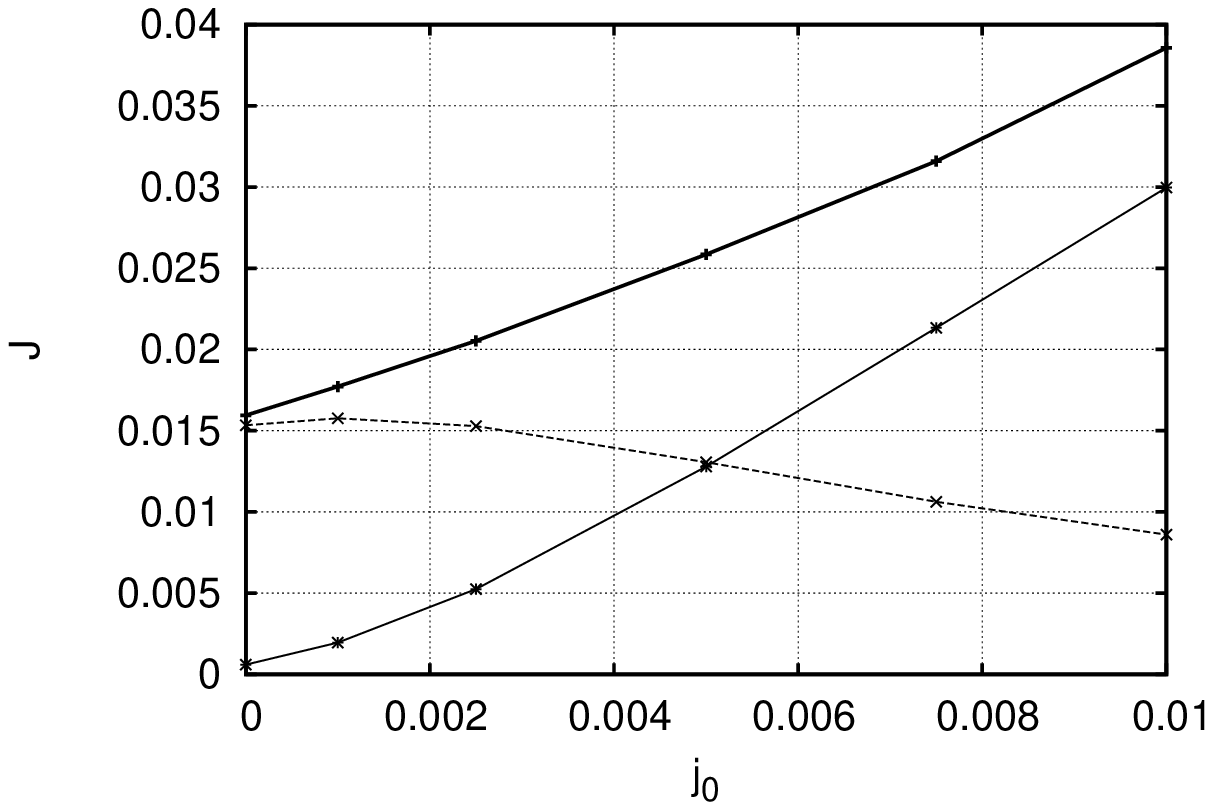}
 \caption{Left panel:
$\hat{J}_{tot}$ (thick solid line),
$\hat{J}_{tot}^{(+)}$ (thin solid line) and 
$\hat{J}_{sur}$ (thick dashed line) are 
plotted against $\hat{j}_0$.
Right panel: $\hat{J}_{tot}^{(+)}$ (thick solid line), 
$\hat{J}_{tot}^{\mu}$ (thin dashed line) and 
$\hat{J}_{tot}^{\kappa}$ (thin solid line) are 
plotted against $\hat{j}_0$.
}
\label{Fig:j0_2}
\end{figure*}

\begin{figure*}
 \includegraphics[scale=0.62]{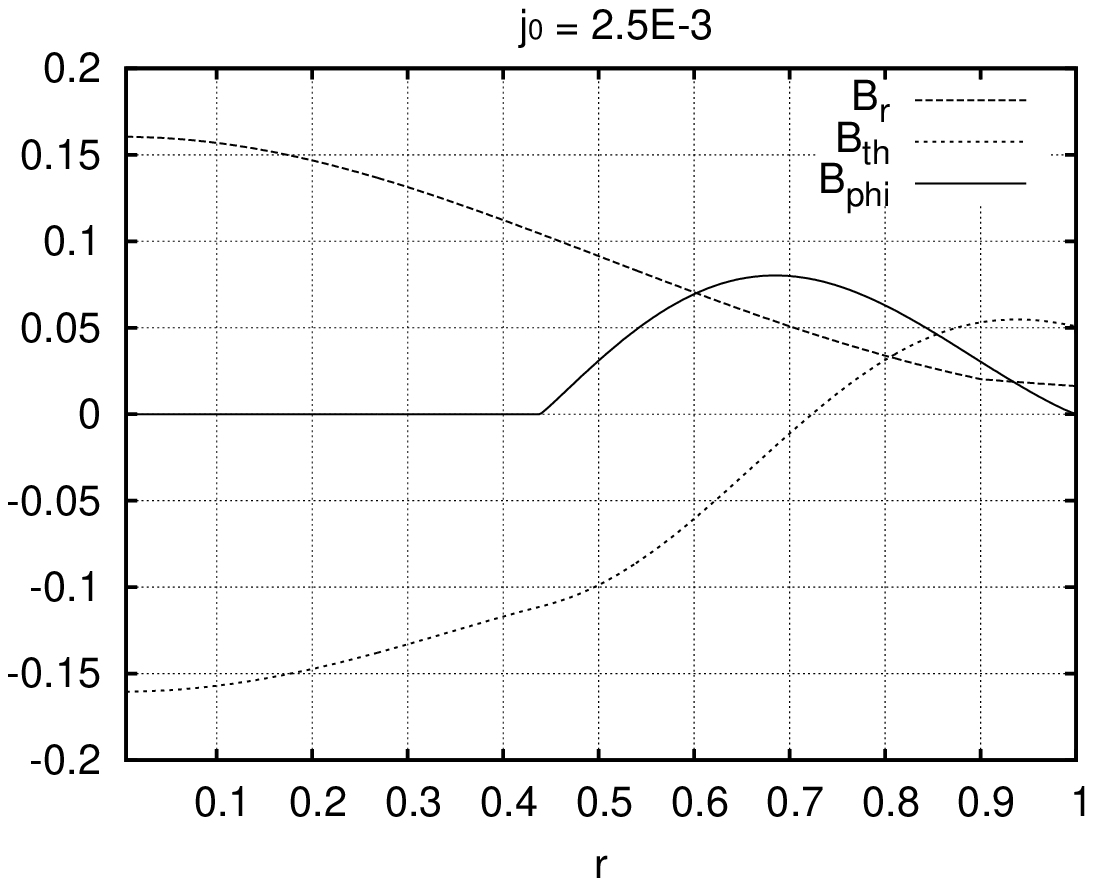}
 \includegraphics[scale=0.62]{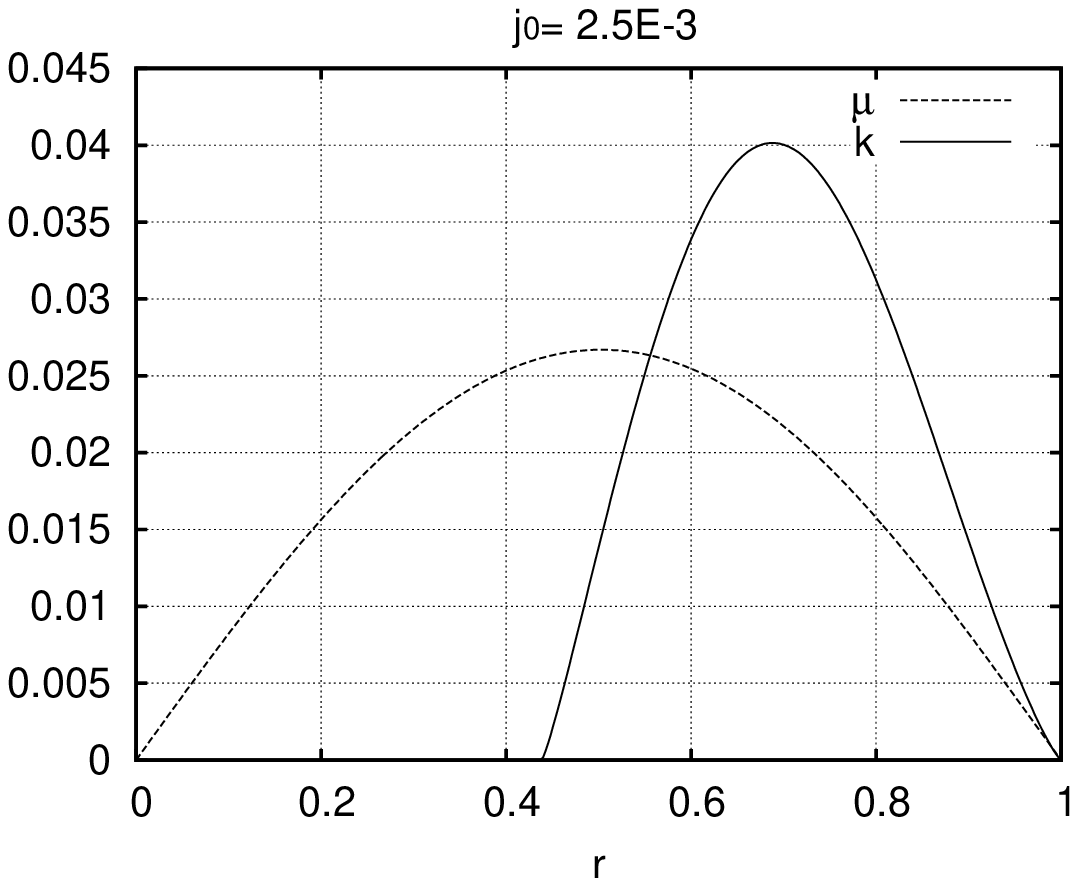}

 \includegraphics[scale=0.62]{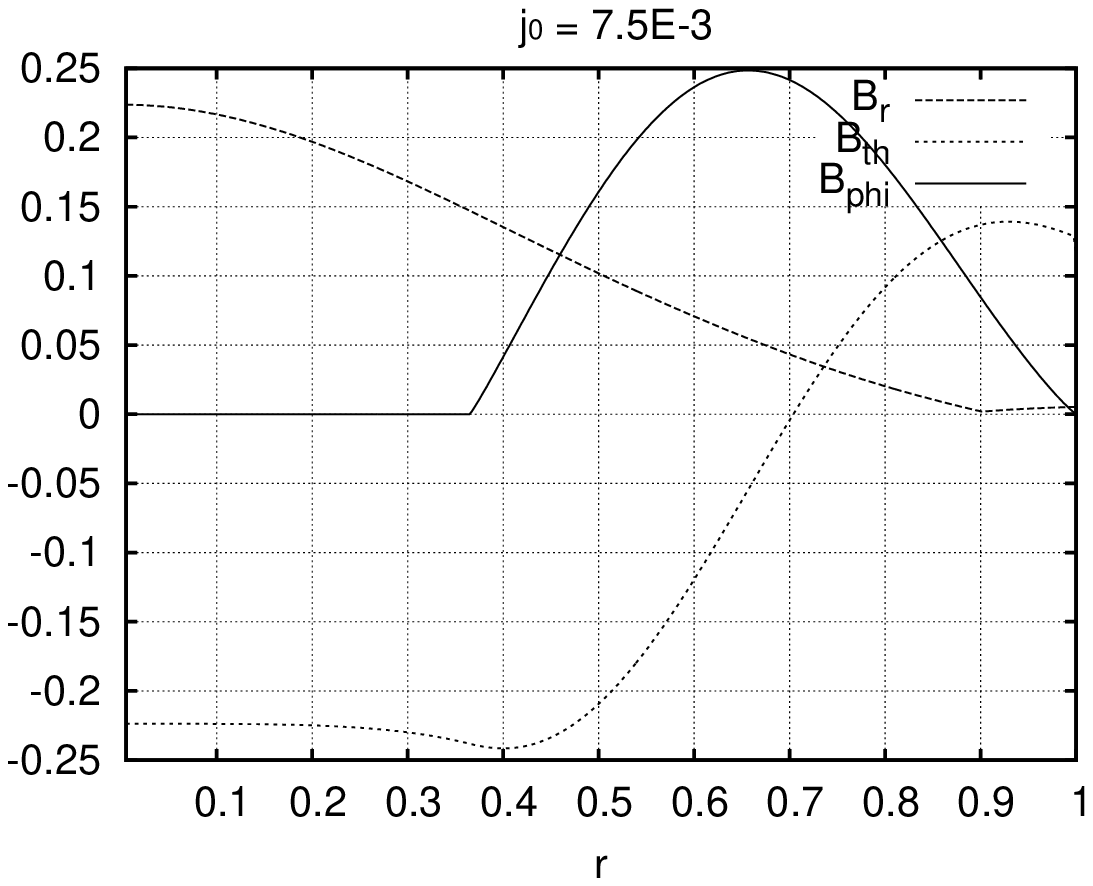}
 \includegraphics[scale=0.62]{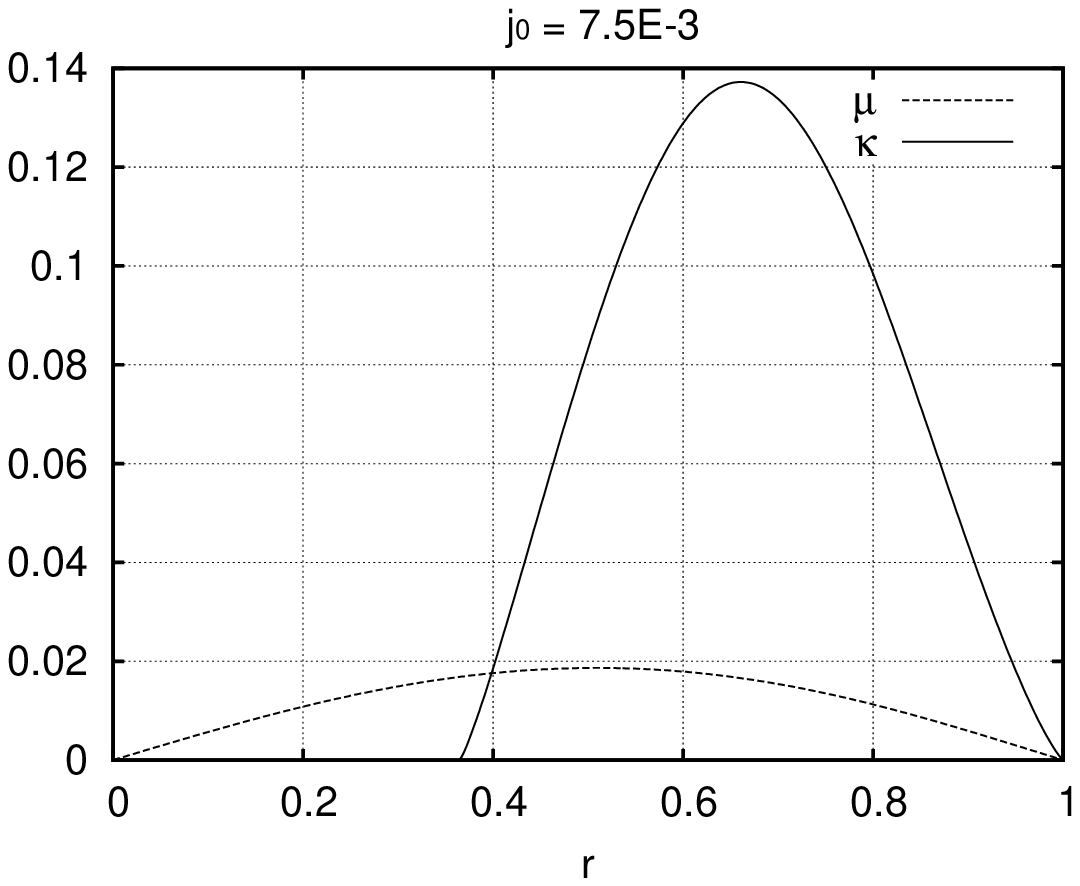}

 \caption{The same as Fig:3 except for 
 configurations with non vanishing surface 
 currents, i.e. configurations with 
 $\hat{j}_{0} = 2.5\times 10^{-3}$ (top) and 
 $\hat{j}_{0} = 7.5\times 10^{-3}$ (bottom).}
\label{Fig:j0_H_r_j_r}
\end{figure*}

As explained in the previous subsection,
in order to exceed the upper bound of 
$\hat{J}_{tot}$ found in this paper
and to reduce the value of $\Psi_{Vmax}$,  
we will try to investigate the magnetized 
stars which contain oppositely flowing 
surface toroidal currents against the 
interior 'bulk' currents which are flowing in a certain 
direction. We will call such oppositely
flowing currents as {\it negative}
currents, hereafter. In short, we will
assume that there could be toroidal 
surface currents which flow to the negative
direction compared to the flow direction
of the interior main currents which we 
will call the {\it interior bulk} currents. In fact the
effects of the presence of the oppositely 
flowing surface currents are similar to 
those of the boundary conditions 
treated in section 5.2 of 
\cite{Glampedakis_Andersson_Lander_2012} 
as we have shown in Sec.\ref{Sec:j_sur_diff}.  
As we shall show, magnetized stars with 
negative surface currents will be able
to sustain much stronger interior bulk currents 
and have much stronger toroidal magnetic 
fields because the oppositely flowing 
surface current cancels the effects
of the interior toroidal currents to 
certain  extent and results in 
configurations which have the following 
two special characteristics.
\begin{enumerate}
 \item In such configurations, although the total 
currents $\hat{J}_{tot}$ are below the
upper bound $\hat{J}_{tot}^{(max)}$ discussed 
before, much stronger positive interior 
bulk currents $\hat{J}_{tot}^{(+)}$ are 
allowed to exist.
\item At the same time, in such 
configurations, the absolute values of the 
magnetic flux function in the outer vacuum 
region can become smaller than those of 
configurations without negatively flowing 
surface currents. Thus the support region
for the arbitrary function $\kappa(\Psi)$
can become wider than that for configurations
without surface currents.
\end{enumerate}

In this subsection, as an example, 
the dipole-like distribution 
for the surface current as Eq.(\ref{Eq:j_sur_dip})
is employed. If the magnetized stars are 
purely spherical with no interior currents 
within the stars, dipole-like surface 
currents result in uniformly distributed 
interior magnetic fields and purely dipole 
exterior magnetic fields (see Fig.\ref{Fig:dipole_quadrupole}). 
Thus, if the surface current densities are 
much stronger than the interior current densities, 
the interior magnetic fields become almost 
uniform and there are no closed magnetic fields 
inside of the stars. The toroidal magnetic 
fields could not appear in such configurations.

We have calculated many stationary configurations 
with surface currents for several values of 
$\hat{\kappa}_0$. As a result, we found that a model 
with $\hat{\kappa}_0 = 10$ has the smallest value 
of ${\cal M}_p / {\cal M}$ in all our stationary
solutions. Thus we will show only configurations
with $\hat{\kappa}_0 = 10$ in this paper, but 
it should be noted that all other models show 
almost the same tendency as that of 
configurations with $\hat{\kappa}_0 = 10$ which
we will describe below. Fig. \ref{Fig:j0} shows 
the values of ${\cal M}_p / {\cal M}$ and 
Fig. \ref{Fig:j0_2} shows the values of $\hat{J}_{tot}$ (thick solid line),
$\hat{J}_{tot}^{(+)}$ (thin solid line) and $\hat{J}_{sur}$ (thick dashed line)
in the left panel and the value of $\hat{J}_{tot}^{(+)}$ (thick solid line),
$\hat{J}_{tot}^{\mu}$ (thin dashed line) and 
$\hat{J}_{tot}^{\kappa}$ (thin solid line) in the right panel
against the values of $\hat{j}_0$ for configurations
with $\hat{\kappa}_0 = 10$. 
In these models, there is no negative current $\hat{j}_\varphi^{(-)}$ in the star.
Therefore, $\hat{J}_{tot}^{(-)} = 0$ and 
$\hat{J}_{tot}^{(+)} = \hat{J}_{tot}^{\kappa} + \hat{J}_{tot}^{\mu}$ in these configurations.
From this figures, we can see that if we increase 
the value of $\hat{j}_0$, the total bulk
current of the magnetized star,
$\hat{J}_{tot}^{(+)}$, becomes larger (Fig.\ref{Fig:j0_2}) and the value of 
${\cal M}_p / {\cal M}$ becomes smaller (Fig.\ref{Fig:j0}). 
We can see from left panel of Fig.\ref{Fig:j0_2}
that the total current $\hat{J}_{tot}$ is almost the same as the upper
bound of the total current $\hat{J}_{tot}^{(max)}$
defined before. However, the total positive current  
$\hat{J}_{tot}^{(+)}$ becomes much larger than this upper bound. 
Especially noted from the right panel in Fig.\ref{Fig:j0_2}, 
the total $\kappa$ current term becomes much larger and
the total $\mu$ current term becomes slightly small.
This can be considered as an evidence that the negative surface current 
cancels some contributions of the positively
flowing interior toroidal current from the $\kappa$ term current. 
It is remarkable that the value of ${\cal M}_p / {\cal M}$ attains about 0.7 
when $\hat{j}_{0} = 7.5\times10^{-3}$. It implies 
that the magnitude of the toroidal magnetic field 
energy is almost the same order as that of the 
poloidal magnetic energy for those models
around $\hat{j}_{0} = 7.5\times10^{-3}$. 
The ratio ${\cal M}_p / {\cal M} \sim 0.65$ seems 
to be the minimum value in the present parameter 
space because we could not succeed in obtaining 
numerical solutions when 
$\hat{j}_0 > 1.0\times10^{-2}$. Since the 
surface current with $\hat{j}_{0} = 2.0 \times 10^{-2}$ 
should be considered tremendously strong,
their fields would become nearly uniform in
the interior and purely dipole in the outside of the star 
by the surface current as seen Fig.\ref{Fig:dipole_quadrupole}. 
Moreover, when there are no closed poloidal 
magnetic field lines inside the stars, 
the magnetized stars cannot have toroidal 
magnetic fields because the poloidal magnetic
fields are originated from the closed current 
densities which are assumed to be parallel to
the closed poloidal magnetic fields as seen
from the current density formula.

Fig. \ref{Fig:j0_H_r_j_r} shows the components 
of the magnetic fields (left) and the 
non-force-free and the force-free term in the 
current density formula (right) are plotted 
against $\hat{r}$. We choose 
$\hat{j}_0 = 2.5 \times 10^{-3}$ (top panels) and 
$\hat{j}_0 = 7.5 \times 10^{-3}$ (bottom panels).
From these panels, we can see both the width 
and the strength of the $\kappa$ term are 
increasing as $\hat{j}_0$, but that the
strength of the $\mu$ term does not change 
so much. This result means that 
the oppositely flowing surface current affects
only the force-free term in the current 
density formula significantly. As discussed before,
since the $\mu$ term is non-force-free
and affected mainly from the global structures
of the stars, i.e. by the value of the axis 
ratio $q$. The distributions of $\mu$ term
in the current density formula would not
show large change. We have computed these
configurations by fixing the values of $q$ and 
$\hat{\Omega}_0$ and the total magnetic 
strengths of the stars are nearly the same 
for the different configurations.
Therefore, the distributions of the $\mu$ terms are 
nearly the same for different values of $\hat{j}_0$.

It is needless to say that in order to 
increase the total current keeping the $\mu$ term 
nearly the same, the $\kappa$ term must become 
larger and stronger as we see in these panels
and Fig.\ref{Fig:j0_2}.
Since the stronger and steeper distributions of 
the $\kappa$ term result in the stronger 
toroidal magnetic fields, the presence of the 
oppositely flowing  surface current should be
required for the larger and stronger 
toroidal magnetic fields.  In other words, 
the oppositely flowing surface current density 
can sustain the strong toroidal magnetic fields. 
The maximum value of the toroidal magnetic field
and the size of the region where the most of 
the toroidal magnetic field exists can be 
increased by adding and increasing the surface 
current densities.
 
\begin{figure*}
 \includegraphics[scale=0.63]{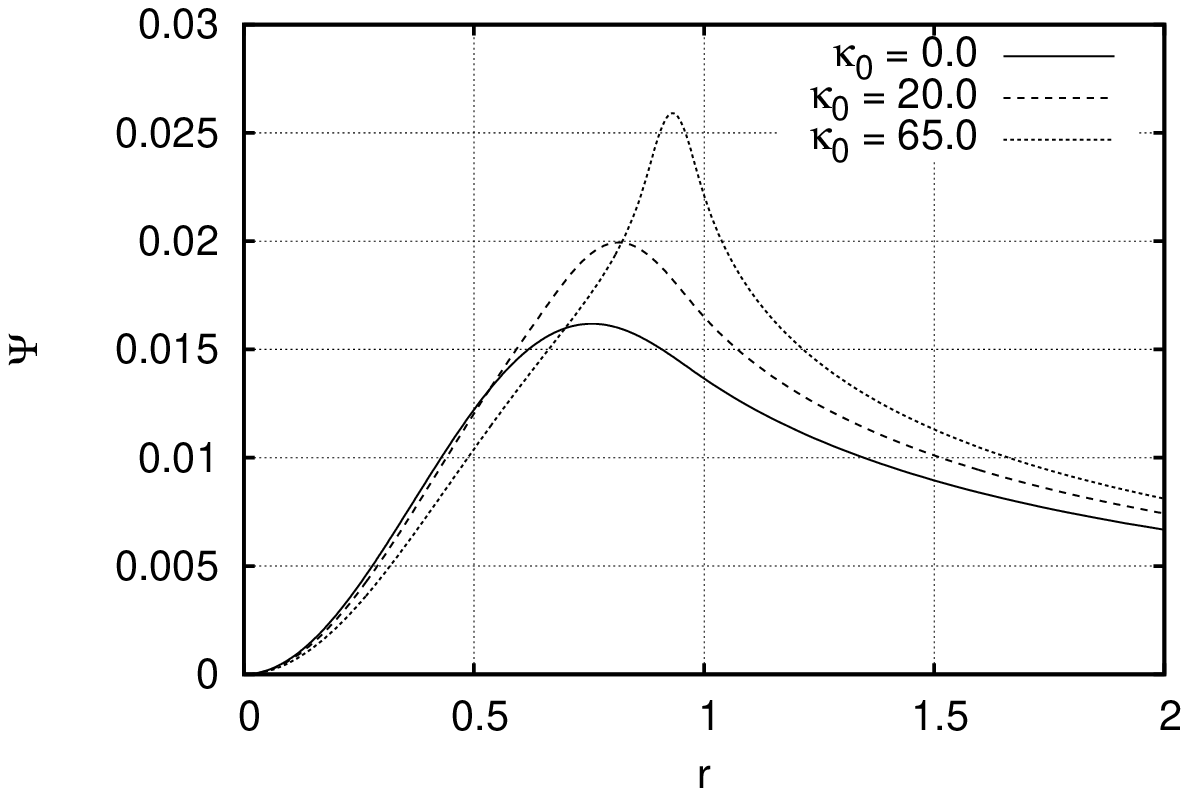}
 \includegraphics[scale=0.63]{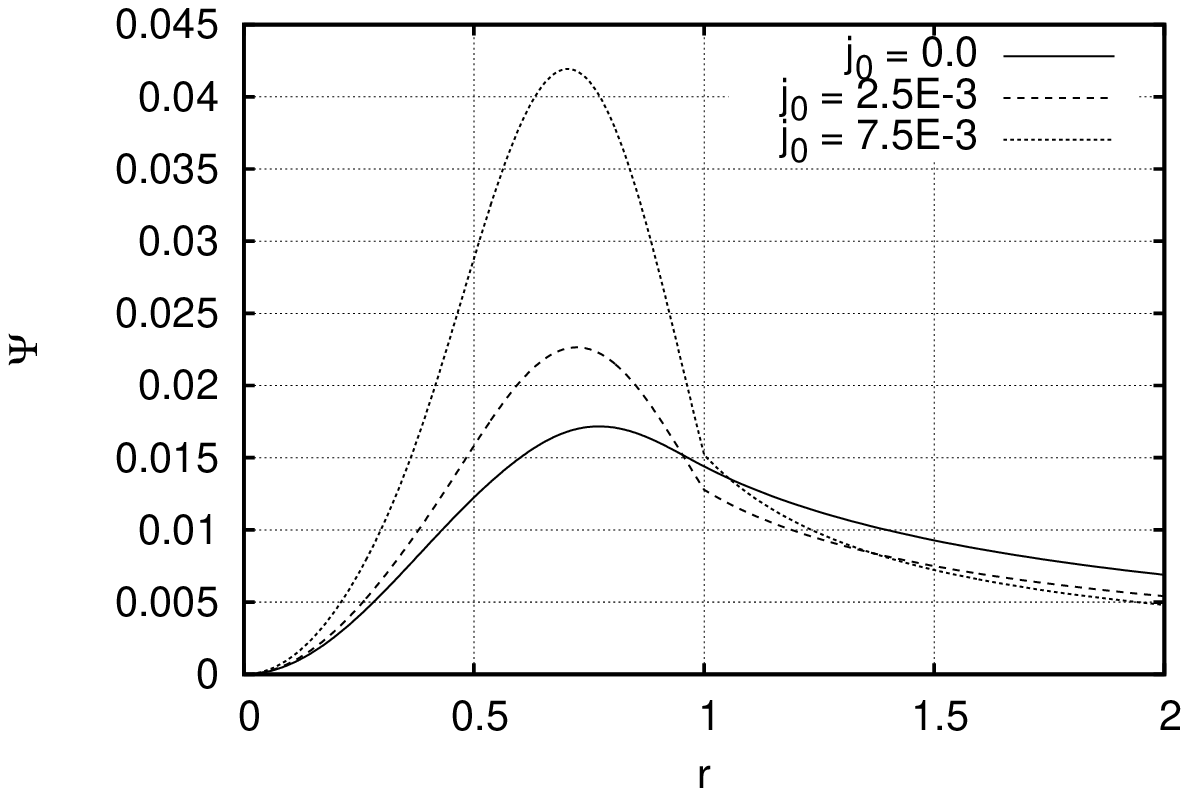}
 \caption{The distributions of magnetic flux function on the equator 
($\theta = \pi/2$).
Left panel: These lines denote $\hat{\kappa}$ = 0.0 (solid line),
$\hat{\kappa}_0$ = 20.0 (dashed line) and $\hat{\kappa}_0 = 65.0$ (dotted line)
model without surface current respectively.
Right panel: These lines denote $\hat{j}_0 = 0.0$ (solid line),
$\hat{j}_0 = 2.5 \times 10^{-3}$ (dashed line) and $\hat{j}_0 = 7.5 \times 10^{-3}$ (dotted line)
model with surface current respectively.}
\label{Fig:Psi_r}
\end{figure*}

In Fig.\ref{Fig:Psi_r} the distributions of the magnetic flux
functions on the equator ($\theta = \pi/2$) for two 
configurations, one without surface current (left panel)
and the other with surface current (right panel), 
are displayed.  As seen from left panel, 
the value of the magnetic flux function at the equatorial
surface for the configuration without surface current 
becomes bigger as the $\hat{\kappa}_0$ increases,
because the maximum value of the flux function becomes bigger.
On the other hand, the value of the magnetic flux at the 
equatorial plane with surface current (right panel) 
does not change very much even if the values of $\Psi_{\max}$ increase.
Therefore, the negative surface current make the 
flux function at the equatorial surface smaller than
that for the configuration without surface current. 
This reduction of the value of the magnetic flux function in 
the vacuum region can allow the wider
support region to exist.

\begin{figure*}
 \includegraphics[scale=0.63]{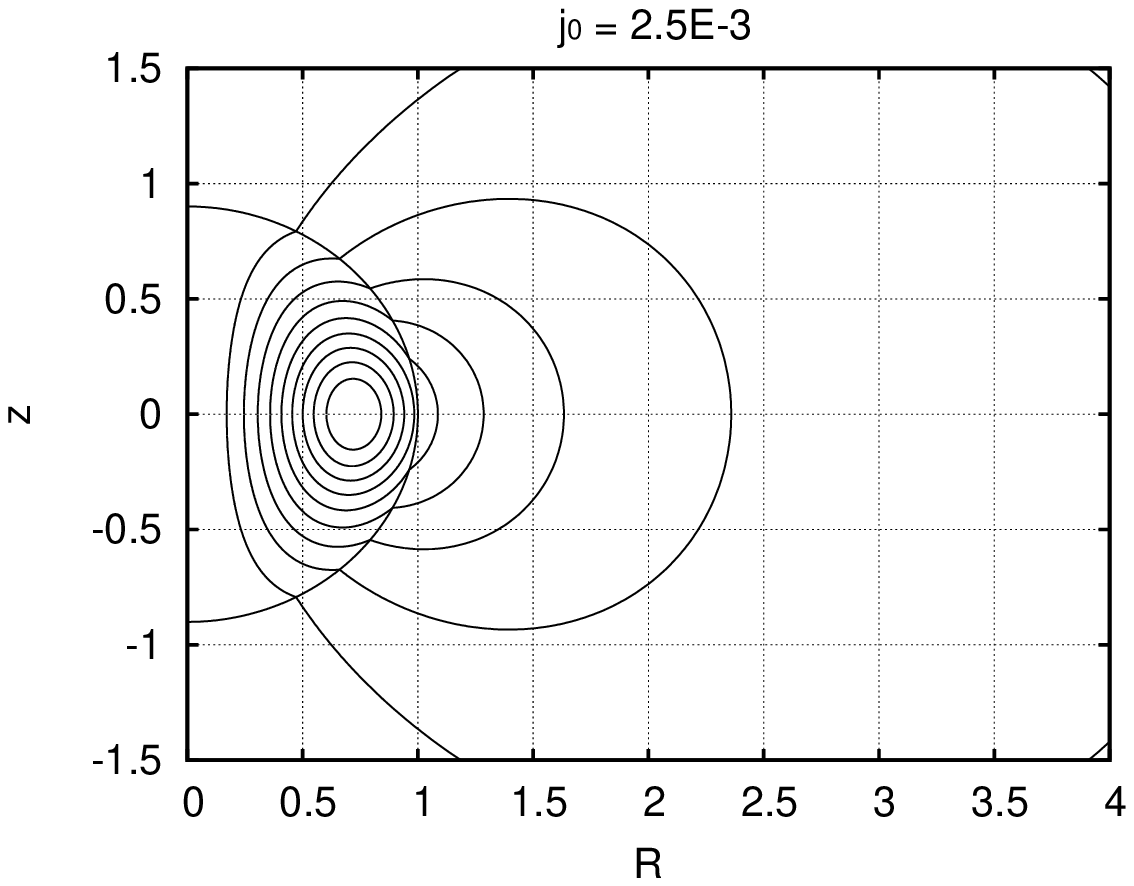}
 \includegraphics[scale=0.63]{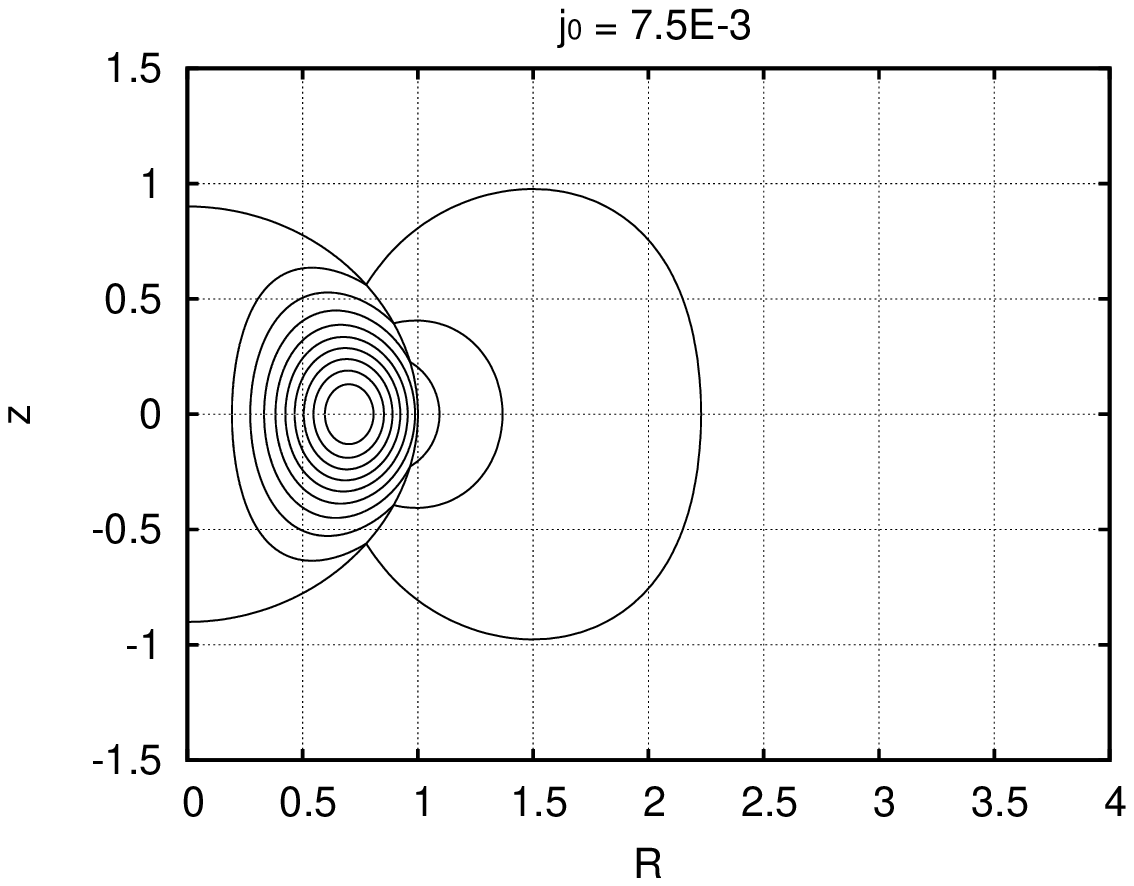}
 \caption{Distributions of the poloidal magnetic 
 fields. Parallel components of the poloidal
 magnetic fields along the surfaces are 
 discontinuous on the stellar surfaces. 
 The values of ${\cal M}_p / {\cal M} 
 \sim 0.8785$ (left) and $\sim 0.6979$ (right).}
 \label{Fig:u_flux}
\end{figure*}

As for the geometry of magnetic fields, 
the surface currents bend the poloidal magnetic 
fields on the stellar surface as we have 
described in Sec.\ref{Sec:j_sur_diff}.
Fig.\ref{Fig:u_flux} shows that the poloidal 
magnetic field lines bend due to the presence of
the surface currents for configurations 
with $\hat{j}_0 = 2.5 \times 10^{-3}$ (left) and 
$\hat{j}_0 = 7.5 \times 10^{-3}$ (right). 

Thus far, nobody has obtained configurations with the 
surface currents in previous works 
(see  Fig.2 in \citealt{Fujisawa_Yoshida_Eriguchi_2012}).
On the other hand, configurations shown
in Fig.\ref{Fig:u_flux} have discontinuities 
on their surfaces. The directions of the 
discontinuities depend on the directions  of 
the surface currents. 
In the northern hemisphere, the outward poloidal 
field lines are bended to the left side by 
the oppositely flowing  surface currents. 
On the other hand, if we add the surface currents 
whose flowing direction is the same as that of the 
interior bulk current, they are bended to the right 
side and the toroidal magnetic fields become weak
because the surface currents flowing to the same
direction as the interior bulk currents work so as to
reduce the strengths of the interior bulk currents.
The discontinuities of the 
magnetic fields in our models are the same as 
those of  \cite{Glampedakis_Andersson_Lander_2012} 
(see Fig.5 in their paper carefully). 
In fact, what they did in their paper,
i.e. by imposing structures in which 
the magnetic fields on the stellar surfaces
have some amount of discontinuities are 
effectively the same as adding the oppositely flowing 
surface currents to the magnetized stars.

\subsection{Configurations with surface 
currents  -- Quadrupole currents}

We consider configurations with other 
surface current distributions.
We add the surface currents expressed 
by Eq.(\ref{Eq:j_sur_quad}), 
which results in the quadrupole 
distribution of the poloidal magnetic 
fields. However, the toroidal magnetic 
field for this surface current cannot 
become large enough.  
As the strength of the surface current
is increased,  we get configurations with
no toroidal magnetic fields whose poloidal 
magnetic fields are not closed inside of 
the stars (see the right panel in 
Fig.\ref{Fig:dipole_quadrupole}).
The toroidal component of the magnetic 
field vanishes in such a configuration.

In order to sustain strong toroidal 
magnetic fields, we need strong toroidal 
currents in the stellar interior as 
discussed for the dipole-like distributions
of the surface currents. However,
as seen from the results for the
quadrupole-like surface currents,
surface currents contain both 
the negative component and the  
positive component in the surface 
currents. Moreover, the strengths of those
oppositely flowing currents are 
the same. Therefore, the total surface 
current due to the purely quadrupole 
surface current of the purely spherical star vanishes as 
\begin{eqnarray}
 \int_{S_{\rm mer}} \hat{j}_{sur} dS = - \int_0^\pi r_s j_0 \sin \theta \cos \theta d \theta = 0.
\end{eqnarray}
Consequently, this kind of surface 
current cannot counteract or cancel the effect
of the interior bulk current density. 
It implies that we need $2^{\ell}$-th 
poles with $ \ell =$ odd numbers 
of the magnetic moments (dipole, 
octopole etc.) or locally strong surface 
currents which are not anti-symmetric 
about the equatorial plane in order 
to sustain the strong toroidal fields.

\section{Reasons for appearance of strong
toroidal magnetic fields --- Coexistence
of oppositely flowing multi-$\varphi$-currents}
\label{Sec:perturb} 

In the previous section we have discussed 
that the presence of negative (oppositely flowing) 
surface currents in addition to positive
interior bulk currents could allow more 
interior currents to exist within the stars. 
In particular, large interior toroidal currents 
could be realized by introducing negatively
flowing surface currents in addition to
positively flowing interior currents. 
Consequently, such negatively flowing toroidal
currents lead to larger positively flowing
total currents, $\hat{J}_{tot}^{(+)}$,  within the stars, 
although the total currents, $\hat{J}_{tot}$,
have their upper bound as explained before.
Thus it is these larger positively flowing 
interior currents, $\hat{J}_{tot}^{(+)}$, that
cause the toroidal magnetic fields stronger. 

Among stationary magnetized stars thus far
obtained in many papers, some authors
have found configurations with large
toroidal magnetic fields by treating the
problem differently 
(see \citealt{Ioka_Sasaki_2004}, 
\citealt{Duez_Mathis_2010}
and \citealt{Yoshida_Kiuchi_Shibata_2012}).
However, no authors have explained reasons
why there can appear such magnetic fields 
with large values of the toroidal magnetic
field energies.

\subsection{Zero-flux-boundary approach: 
$N = 0$ magnetized spherical configurations}

In this subsection, we will reconsider simple 
configurations with large toroidal 
magnetic fields from our standpoint of 
taking the negatively flowing currents into
account. In order to clarify the reasons 
for existence of large toroidal magnetic fields,
it would be helpful to employ as simple 
configurations as possible. 

As for the mechanical structures of the magnetized 
stars we consider $N = 0$ polytropes, 
i.e, incompressible fluids. Although the magnetic 
fields might become very strong, the shapes of 
the stellar configurations are assumed to be spheres. 
It is possible to consider that strongly magnetized 
stars have spherical surfaces by confining 
all the magnetic fields within the surfaces.
This situation could be realized by treating
the closed magnetic fields not only for the toroidal
fields but also for the poloidal fields.
Under these assumptions, the formulations 
used  by several authors 
(\citealt{Chandrasekhar_Prendergast_1956};
\citealt{Prendergast_1956};
\citealt{Duez_Mathis_2010}; 
\citealt{Glampedakis_Andersson_Lander_2012}; 
\citealt{Yoshida_Kiuchi_Shibata_2012}) 
can be applied. Since our purpose of this part
of the paper is not to obtain 'new' stationary
configurations but to understand the {\it reasons}
for appearance of large toroidal magnetic fields,
we will follow mostly the zero-flux-boundary 
scheme of \cite{Prendergast_1956} (see also
\cite{Duez_Mathis_2010}), but we take account
of the presence of negatively flowing surface
currents as well as negatively flowing interior 
currents in addition to
positively flowing interior currents.

As for the arbitrary functions, we choose 
the functional form of $\kappa$ as follows:
\begin{eqnarray}
\kappa(\Psi) \equiv  \kappa_0 \Psi,
\end{eqnarray}
and consequently 
\begin{eqnarray}
\D{\kappa(\Psi)}{\Psi} \equiv  \kappa_0.
\end{eqnarray}
This choice for $\kappa(\Psi)$ as well as the form
for $\mu(\Psi) = \mu_0 (\mbox{constant})$ have 
been commonly used in almost all 
zero-flux-boundary approaches which have 
treated
configurations with closed dipole magnetic fields 
(see \citealt{Prendergast_1956},
\citealt{Ioka_Sasaki_2004}, 
\citealt{Duez_Mathis_2010}
and \citealt{Yoshida_Kiuchi_Shibata_2012}).
From Eq.(\ref{Eq:current}), the form 
of the current density becomes as below:
\begin{eqnarray}
 \frac{j_\varphi}{c} = \frac{\kappa_0^2}{4\pi r \sin \theta} \Psi + \rho r \sin \theta \mu_0.
 \label{Eq:jphi_closed}
\end{eqnarray}

Since the GS equation is an elliptic type 
partial differential equation of the
second order, we need to impose
boundary conditions to obtain solutions
consistently. We should note that, in all 
zero-flux-boundary approaches thus far carried out, 
the constant $\kappa_0$ plays a role as an 
eigenvalue of the problem because boundary 
conditions have been imposed at finite places in 
the space in most investigations. 
One example of the boundary conditions
might be as follows:
\begin{eqnarray}
\Psi |_{r=0} & = & 0 \ , 
(\mbox{at the centre}) , 
 \label{Eq:boundary_closed0}  \\
\Psi |_{r=r_s^{(s)}} & = &0  \ , 
(\mbox{on the stellar surface}) \ ,
 \label{Eq:boundary_closed}
\end{eqnarray}
where $r_s^{(s)}$ is the radius of the 
spherical incompressible magnetized stars 
treated in this section. 
It should be noted that solutions with 
$d\Psi / dr|_{r=r_s^{(s)}} = 0$ in our models,
which can be found only after we have obtained 
stationary configurations and checked 
values of the derivative $
d\Psi / dr|_{r=r_s^{(s)}}$ for all the models,
are essentially the same as those of 
\cite{Duez_Mathis_2010}. It should be stressed 
once again that solutions which satisfy the 
condition $d\Psi / dr|_{r=r_s^{(s)}} = 0$ 
would not always be found. It would be
fortunate if one could find such solutions  
not by imposing that condition as one of 
boundary conditions but by just calculating 
solutions with the boundary conditions 
(\ref{Eq:boundary_closed0}) and
(\ref{Eq:boundary_closed}).

\subsection{Magnetic flux functions for 
spherical incompressible fluids
with magnetic fields confined within
the stellar surfaces}
\label{Sec:closed_fields_model}

In this section, we continue to follow mostly the 
formulation of \cite{Prendergast_1956} 
(see also \citealt{Duez_Mathis_2010}), in which 
the surface currents were not taken into
account explicitly, but in this paper we 
include the surface currents explicitly 
by modifying their formulation.

They treated incompressible fluid stars,
i.e. $N = 0$ polytropes
by specifying arbitrary functions
as we have already explained before.
Although incompressible stars seem far from
realistic situations, from the
standpoint of considering oppositely flowing 
currents including surface currents stressed
in this paper, 
it is very useful to be able to get such 
solutions and discuss the role of 
the oppositely flowing currents analytically. 
Nevertheless, in this paper, we will also 
compute $N=1$ polytropes numerically and 
discuss the effect of the equation of state.

For the functional forms we have chosen,
the Grad-Shafranov equation, Eq.(\ref{Eq:GS}), 
can be written as below:
\begin{eqnarray}
\begin{split}
\Delta^* \Psi + \kappa_0^2 \Psi  = - 4 \pi \mu_0 \bar{\rho} r^2 \sin^2 \theta \equiv S(r,\theta).
\end{split} 
\end{eqnarray}
where $\bar{\rho}$ is the averaged value
of the density. It should be noted that this is 
a linear equation for the magnetic flux function 
$\Psi$. When $\rho$ is constant throughout the
stellar interior, we can integrate this
GS equation easily by expressing the solution
in the integral form by using Green's function 
and spherical Bessel functions and 
Gegenbauer polynomials
as follows:
\begin{eqnarray}
 \Psi (r, \theta) = \Psi_s + \Psi_h,
\end{eqnarray}
\begin{eqnarray}
\begin{split}
& \Psi_s \equiv \sum_{\ell=0} \Psi_{\ell} \\
& =   - 4\pi \mu_0 \sum_{\ell=0} \kappa_{0,\ell} \left[\frac{2\ell + 3}{2(\ell+1)(\ell+2)}\right] 
\sin^2 \theta C_\ell^{3/2} (\cos \theta)  \\
& \Bigg\{ \frac{r}{r_s^{(s)}} J_{\ell+1} \left(\kappa_{0,\ell} \frac{r}{r_s^{(s)}} \right) \int_r^{r_s^{(s)}} r' Y_{\ell+1}
\left(\kappa_{0,\ell} \frac{r'}{r_s^{(s)}} \right) \bar{\rho}  \, dr' 
\nonumber \\
 &+ \frac{r}{r_s^{(s)}} Y_{\ell+1} \left(\kappa_{0,\ell} \frac{r}{r_s^{(s)}} \right) \int_0^{r} r' J_{\ell+1}
\left(\kappa_{0,\ell} \frac{r'}{r_s^{(s)}} \right) \bar{\rho}  \, dr'
\Bigg\} \\ &  \int_{-1}^{1} S(r', \cos\theta') C^{3/2}_\ell (\cos \theta') d \cos \theta', 
\end{split}
\end{eqnarray}
\begin{eqnarray}
\begin{split}
 &\Psi_h = \sum_{\ell=0} \Bigg[ K_{1,\ell} \kappa_{0,\ell} \frac{r}{r_s^{(s)}} J_{\ell+1}
\left(\kappa_{0,\ell} \frac{r}{r_s^{(s)}}\right) \nonumber \\
&+ K_{2,\ell} \kappa_{0,\ell} \frac{r}{r_s^{(s)}} Y_{\ell+1}
\left(\kappa_{0,\ell} \frac{r}{r_s^{'s)}}\right)
\Bigg] \sin^2 \theta C_\ell^{3/2} (\cos\theta).
\end{split}
\end{eqnarray}
Here $\Psi_s$ and $\Psi_h$ denote the inhomogeneous
solution and the homogeneous solution to the GS
equation, respectively and $J_\ell$'s and 
$Y_\ell$'s are the spherical Bessel functions 
of the first kind and the second kind, respectively 
and $C_\ell^{3/2}$'s are the Gegenbauer polynomials.
$\Psi_{\ell}$, $\kappa_{0,\ell}$ and $K_{1,\ell}$, 
$K_{2,\ell}$ denote the $\ell$-th 
expansion terms of the magnetic flux function, 
the eigenvalues corresponding to $\kappa$
for the $\ell$-th component equations
appearing in the expansion of the magnetic flux 
function as above, and coefficients of 
homogeneous solutions, respectively. It should 
be noted that, here in this paper, we consider 
only the dipole term $(\ell = 0)$ which can be
considered as the simplest configuration 
for the spherical incompressible 
magnetized fluid star. Moreover it is
important to note that even such simple 
configurations contain the essential natures 
of the configurations we are seeking to understand. 

As we have described before,  we impose boundary conditions for $\Psi$.
One condition is  $\Psi = 0$ at the 
centre of the star. This condition can be 
fulfilled simply in our situation here 
by setting $K_2^0 = 0$,  because the spherical Bessel function 
of the second kind $Y_{\ell=0}$ does not vanish 
at the centre ($r=0$) for the homogeneous 
solutions. As a result, we obtain the general 
expression of the solution as follows:
\begin{eqnarray}
 \begin{split}
&  \Psi = K_1 \kappa_0 \frac{r}{r_s^{(s)}} J_1 \left(\kappa_0 \frac{r}{r_s^{(s)}}\right) \sin^2 \theta  \\
  &- 4\pi \mu_0 \kappa_0 \sin^2 \theta \Bigg\{ 
\frac{r}{r_s^{(s)}} J_{1} \left(\kappa_0 \frac{r}{r_s^{(s)}} \right) \int_r^{r_s^{(s)}} Y_{1}
\left(\kappa_0 \frac{r'}{r_s^{(s)}} \right) \bar{\rho} r'^3 \, dr' 
\nonumber \\
& + \frac{r}{r_s^{(s)}} Y_{1} \left(\kappa_0 \frac{r}{r_s^{(s)}} \right) \int_0^{r} J_{1}
\left(\kappa_0 \frac{r'}{r_s^{(s)}} \right) \bar{\rho} r'^3 \, dr'
\Bigg\}.
 \end{split}
\end{eqnarray}
Explicit forms of $J_1$, $Y_1$ and 
$C_0^{3/2}$ are as below:
\begin{eqnarray}
 J_1(\lambda) = \frac{1}{\lambda^2} (\sin \lambda - \lambda \cos \lambda),
\end{eqnarray}
\begin{eqnarray}
 Y_1(\lambda) = - \frac{1}{\lambda^2} (\cos \lambda + \lambda \sin \lambda),
\end{eqnarray}
\begin{eqnarray}
 C_0^{3/2}(\cos \theta) = 1.
\end{eqnarray}
We denote $\kappa_0^0$ as $\kappa_0$ and $K_1^0$ 
as  $K_1 $ for simplicity in 
the following part of this paper.

Next, we impose the other boundary condition 
$\Psi = 0$ at the stellar surface.
This condition is written as follows
\begin{eqnarray}
\begin{split}
  K_1 J_1 \left(\kappa_0 \right)  
  = 4\pi \mu_0
 Y_{1} \left(\kappa_0 \right) \int_0^{r_s^{(s)}} J_{1}
\left(\kappa_0 \frac{r'}{r_s^{(s)}} \right) \bar{\rho} r'^3 \, dr' .
\end{split}
\label{Eq:K_1}
\end{eqnarray}
From this equation we can obtain a relation
between $\kappa_0$ and $K_1$ of our problem at hand.
Thus just by giving either $\kappa_0$ or $K_1$,
one {\it complete} solution in our problem 
can be obtained. This is a nice
feature of the simplest configurations, i.e.
the $N = 0$ polytropic configurations 
only with the $\ell = 0$ component for
the confined poloidal closed magnetic fields.

Finally, we will derive the surface current for
our problem. The homogeneous term of this 
solution is related to the surface current 
as we have calculated in Sec.\ref{Sec:j_sur_int}. 
Thus the surface current is expressed as 
\begin{eqnarray}
  \frac{j_{sur} (\theta)}{c} = \frac{1}{4\pi} (B_\theta^{ex} - B_\theta^{in}  ) = 
\frac{1}{4\pi r \sin \theta} \P{\Psi (r, \theta)}{r}\Bigg|_{r=r_s^{(s)}}.
\end{eqnarray}

Since the solution for the magnetic flux function
behaves as $\sin^2 \theta$, the following
quantity becomes a constant and so
we will write its constant value as $j_0$:
\begin{eqnarray}
 \frac{1}{4 \pi r \sin^2 \theta}\P{\Psi (r, \theta)}{r}\Bigg|_{r=r_s^{(s)}} 
  \equiv j_0.
\end{eqnarray}
Thus the distribution of the surface current can be
written as below:
\begin{eqnarray}
 \frac{j_{sur}}{c} = j_0 \sin \theta.
\end{eqnarray}
Explicit forms of $K_1$ and 
$j_0$ can be found in Appendix \ref{App:K_1}.

\begin{figure*}
 \includegraphics[scale=0.6]{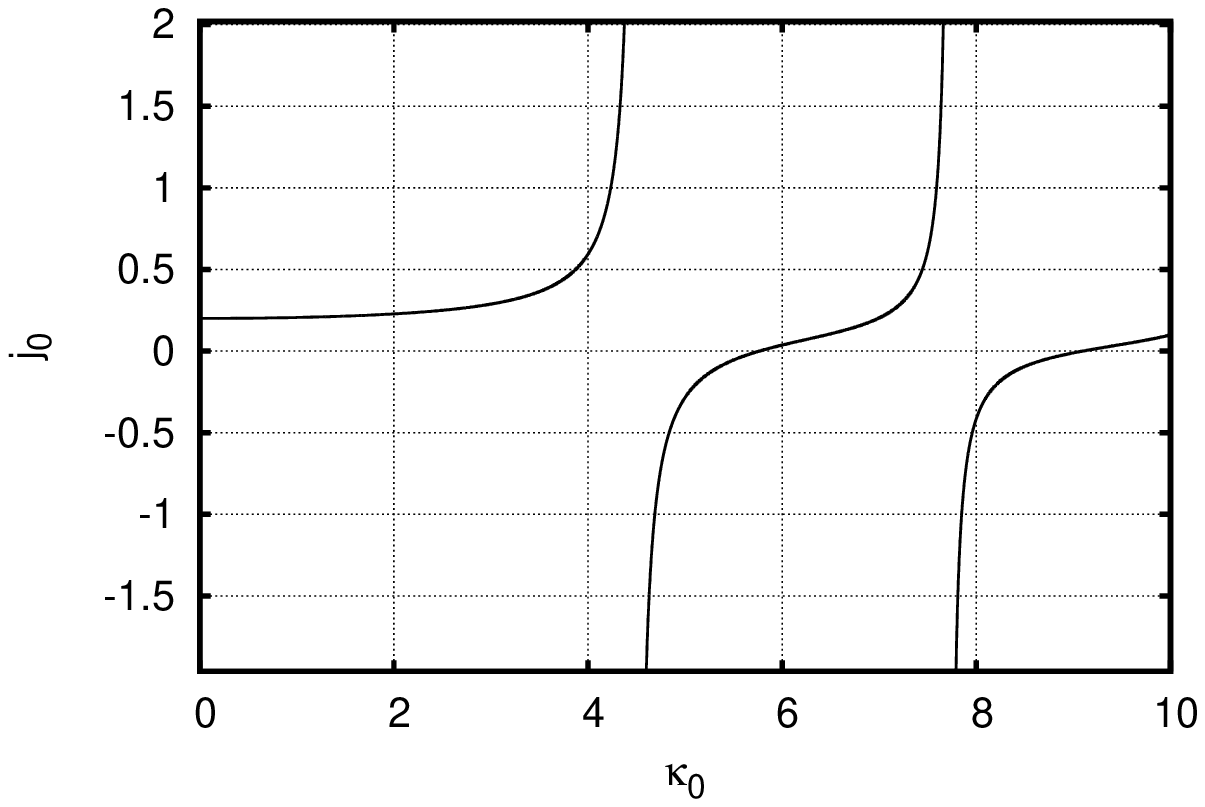}
 \includegraphics[scale=0.6]{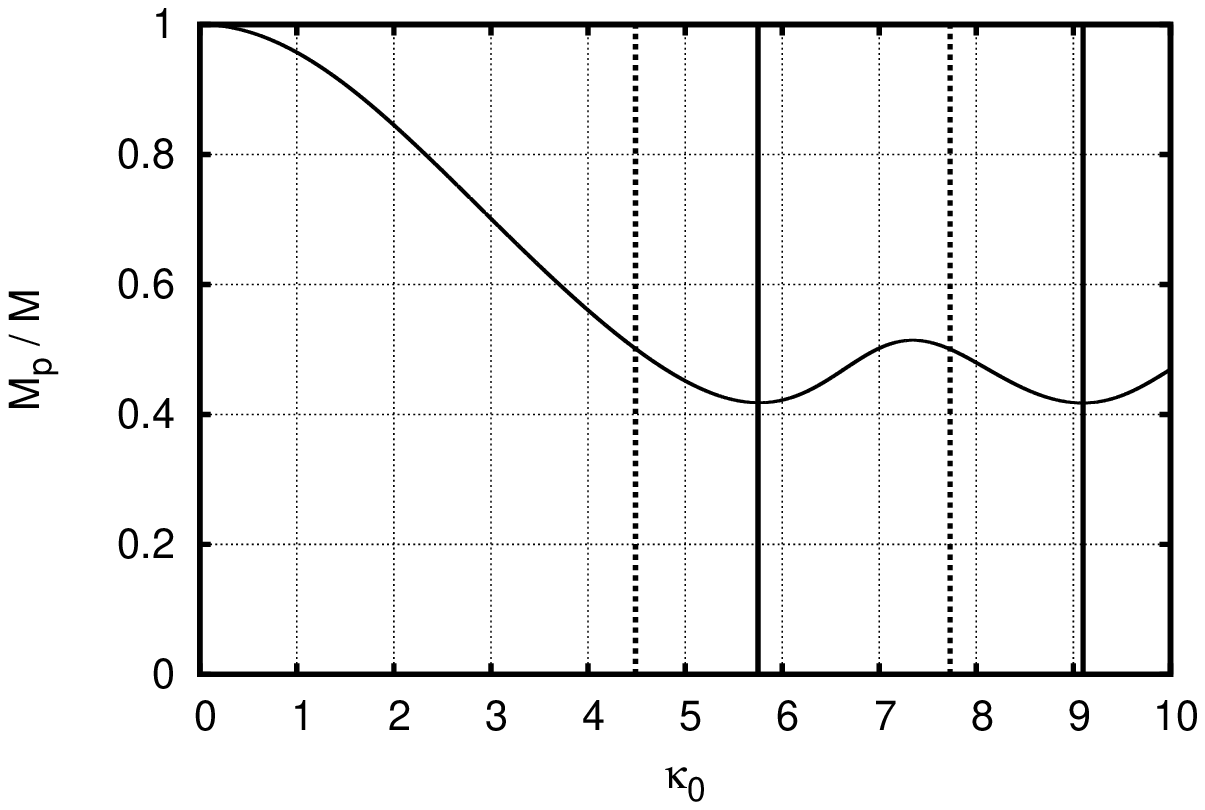}
 \caption{Left: Plotted is  $j_0$ against 
$\kappa_0$ for configurations with the parameter 
satisfying $r_s^{(s)2} \mu_0 \bar{\rho} = -1$.
We can see that there are {\it two solutions
without surface currents} 
at $\kappa_0 \sim$ 5.76 and at 
$\kappa_0 \sim$ 
9.10 and that there exist two singularities 
at $\kappa_0 \sim$ 4.49 and at
$\kappa_0 \sim$ 7.73 within the range $0. \le 
 \kappa_0 \le 10.$.
Right: The energy ratio ${\cal M}_p / {\cal M}$ 
is plotted against $\kappa_0$.
The solid lines and dotted lines denote
the values of $\kappa_0$ for the 
solutions without surface currents 
and the singularities, respectively. 
The toroidal magnetic field vanishes 
at $\kappa_0 = 0$. }
\label{fig:j_0_kappa}
\end{figure*}
Therefore, if the star has negative surface 
currents, values of the magnetic flux functions 
in the large part of the stellar interiors
are positive because of 
$\partial \Psi / \partial r < 0$ on the 
stellar surfaces. 

\begin{figure*}
 \includegraphics[scale=0.62]{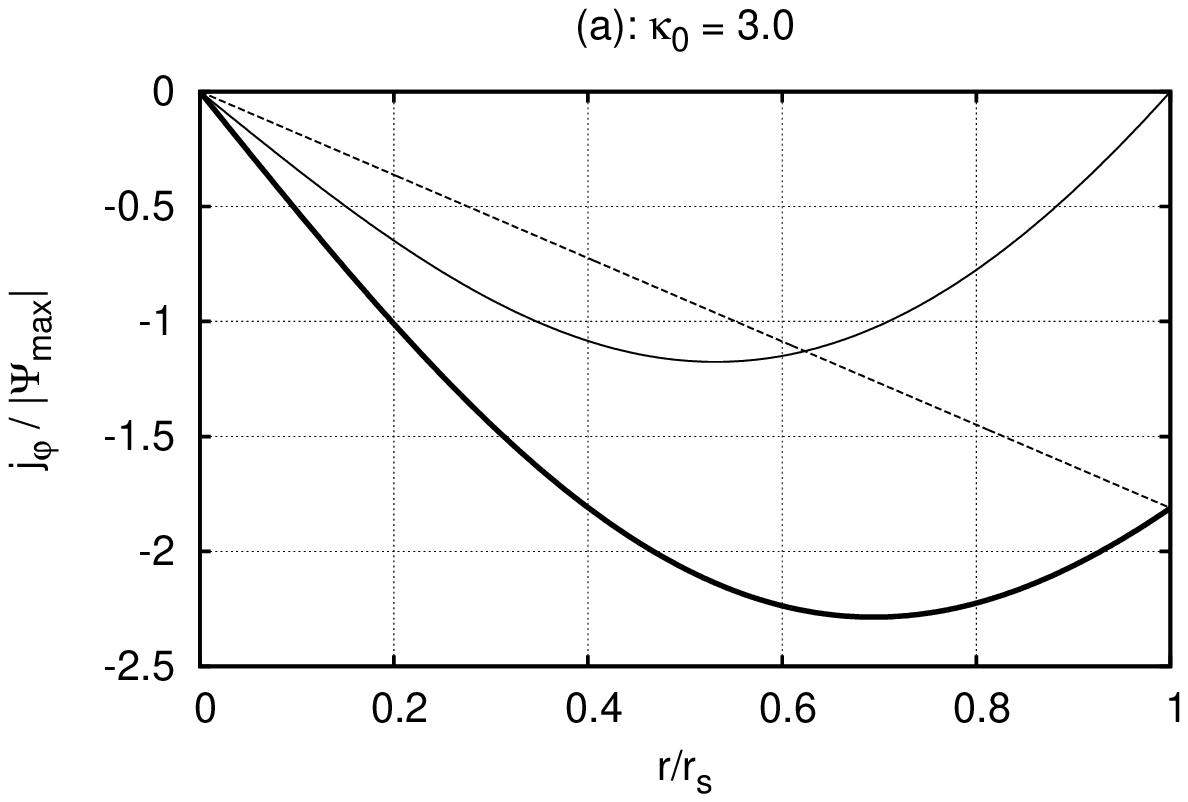}
 \includegraphics[scale=0.62]{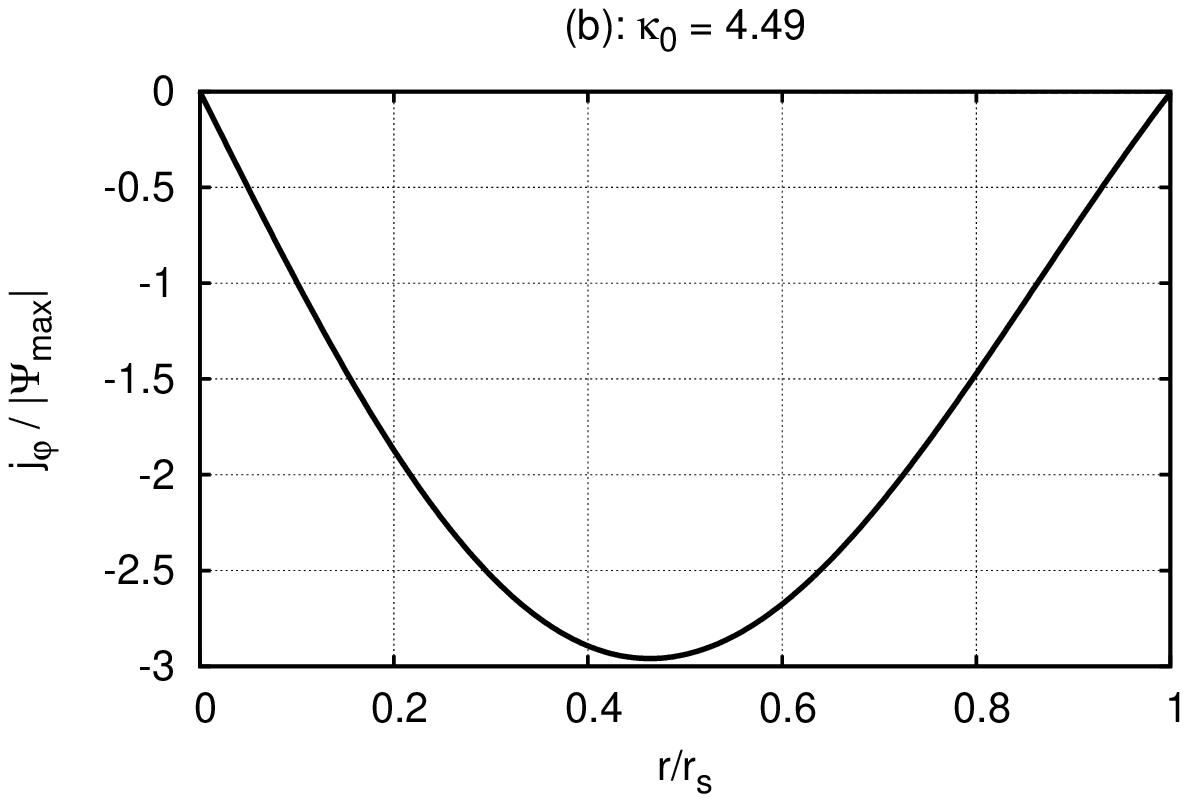}
 \includegraphics[scale=0.62]{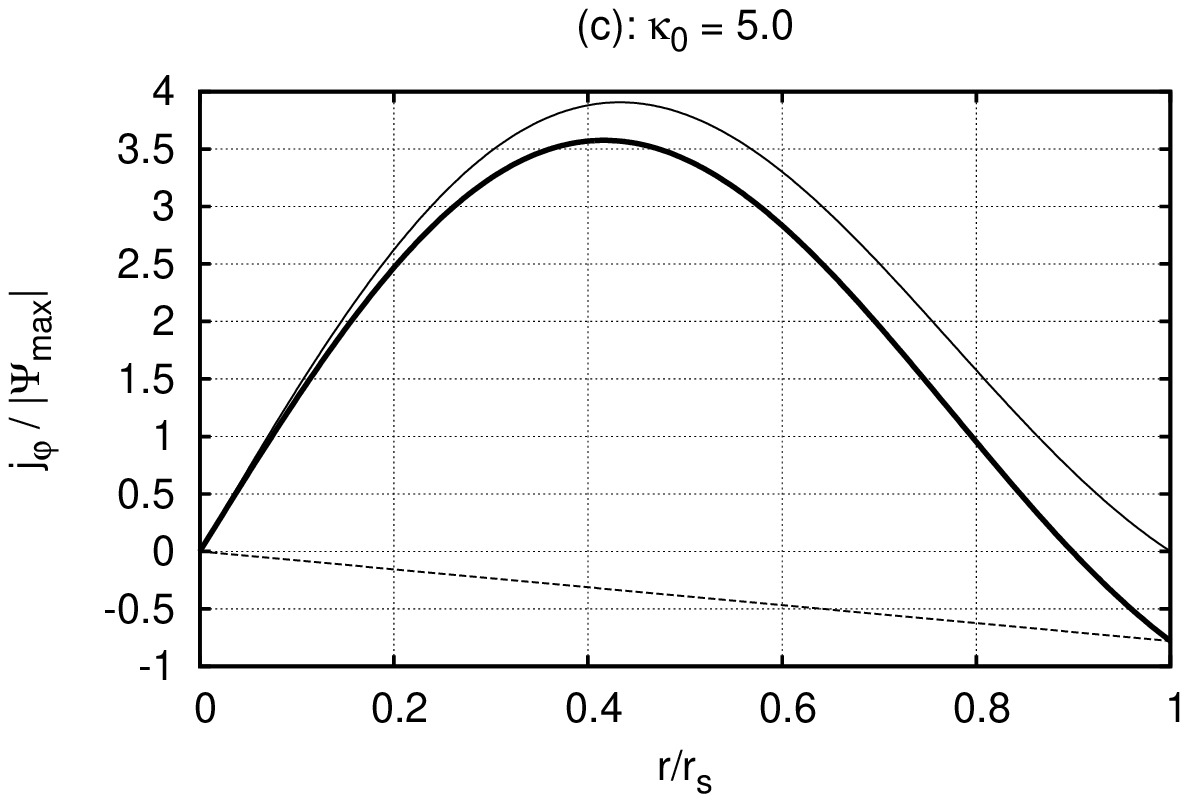}
 \includegraphics[scale=0.62]{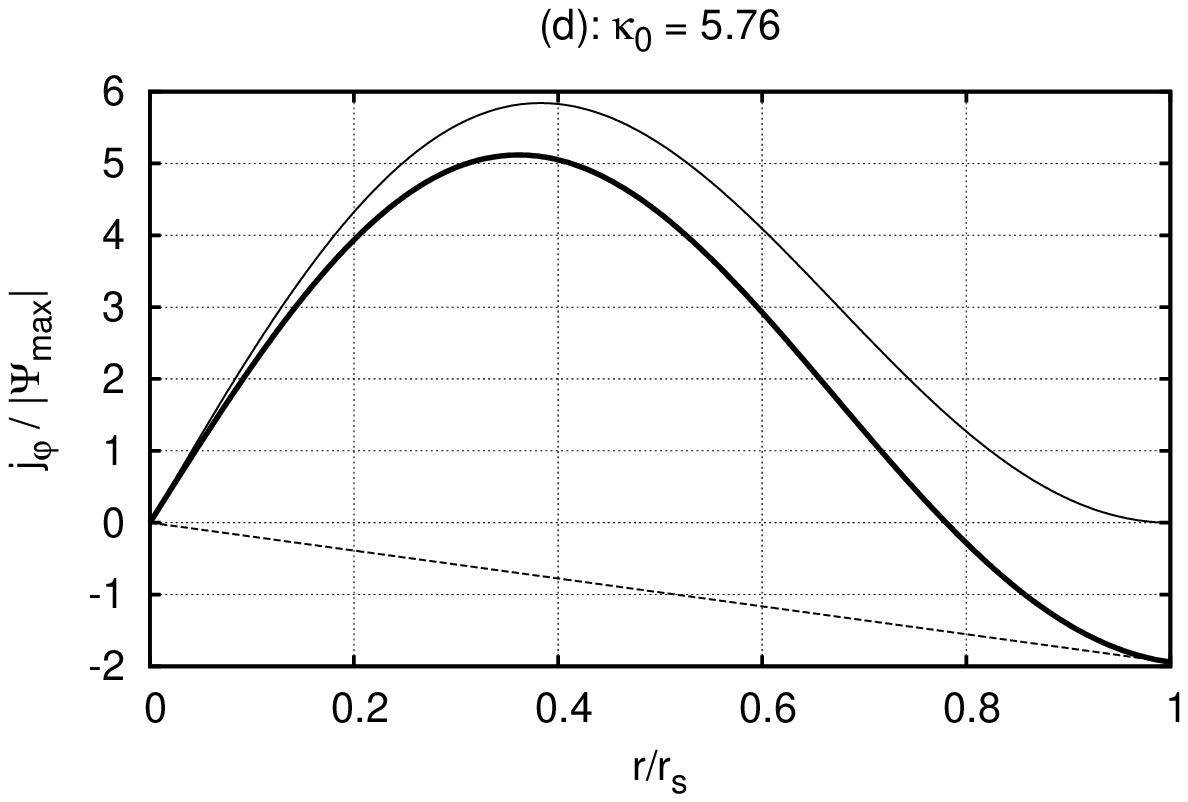}
 \caption{The distributions of $j_\varphi/c$ normalized 
 by the maximum strength of $|\Psi_{\max}|$, i.e. 
 the absolute value of the maximum of the
 flux function, along
 the normalized radius on the equatorial plane. 
 Each line denotes 
 the value of $j_\varphi$ (thick solid line), 
 the component of $j_\varphi$ due to the $\kappa$ 
 term (thin solid line) and 
 the component of $j_\varphi$ due to the $\mu$ 
 term (thin dotted line). 
 Type (a): In the range of $0 < \kappa_0 < \kappa^1_s$
      the $\varphi$ component of the current is 
      negative in the whole star
      and the contributions to the current from 
      the $\kappa$ term and the $\mu$ term are 
      also negative. 
      ${\cal M}_p$ / ${\cal M}$ = 0.701 model.
 Type (b): In the range of $\kappa_0 \sim \kappa_s^1$, 
      the contribution to the $\varphi$ component of 
      the current due to 
      the $\mu$ term is nearly zero because of the 
      large contribution from the $\kappa$ term. 
      ${\cal M}_p $ / ${\cal M}$ = 0.501 model.
 Type (c): In the range of $\kappa_s^1 < \kappa_0 < 
      \kappa_m^1$, the $\varphi$ component of the 
      current is positive in
      most of the stellar interior.  
      The contribution to the current due to 
      the $\kappa$ term is positive.
      whereas the contribution to the current due 
      to the $\mu$ term is negative.
      ${\cal M}_p$ / ${\cal M}$ = 0.452 model. 
  Type (d): At $\kappa_0 = \kappa_m^1$,
      the surface current vanishes because $j_0$
      becomes zero, It is remarkable that
      not only the component of the current 
      due to the $\kappa$ term but also its 
      derivative with respect to the position
      become 0 at the stellar surface. It
      corresponds to the null surface current. 
      ${\cal M}_p$ / ${\cal M}$ = 0.417 model.
 }
 \label{fig:j_j_max}
\end{figure*}

\begin{figure*}
 \includegraphics[scale=0.63]{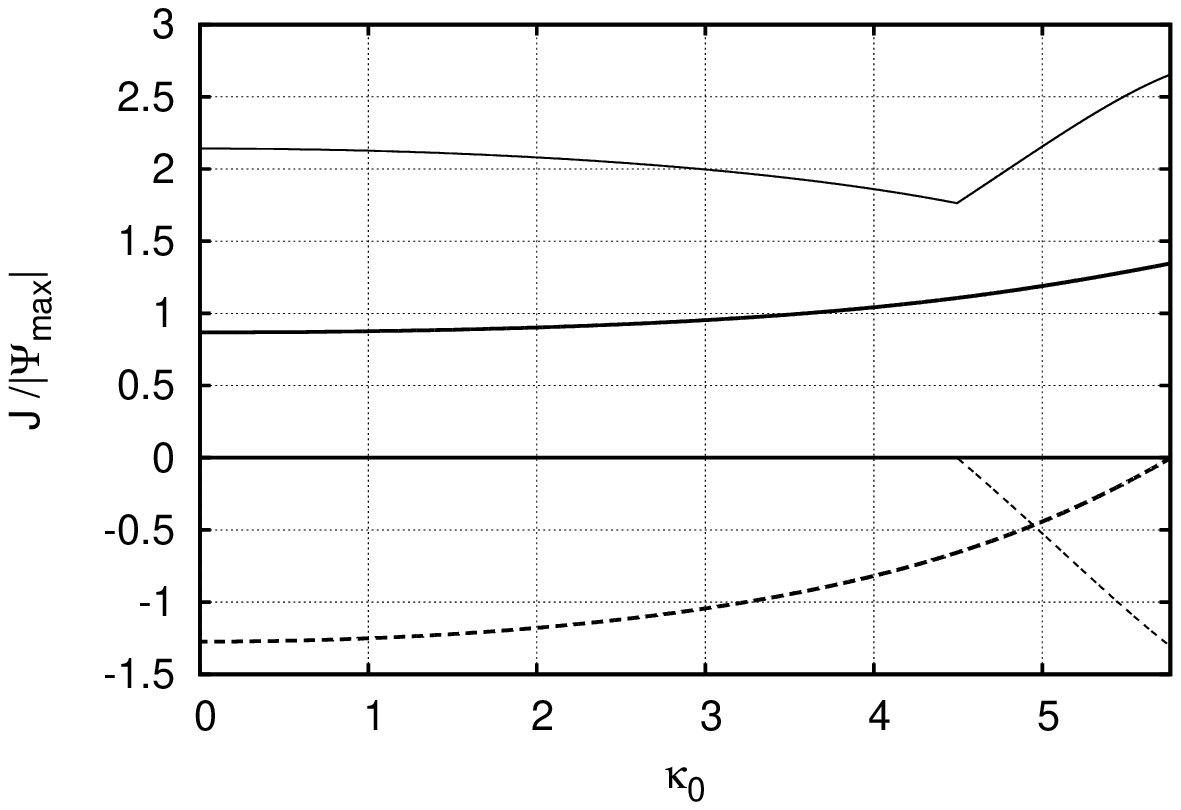}
 \includegraphics[scale=0.63]{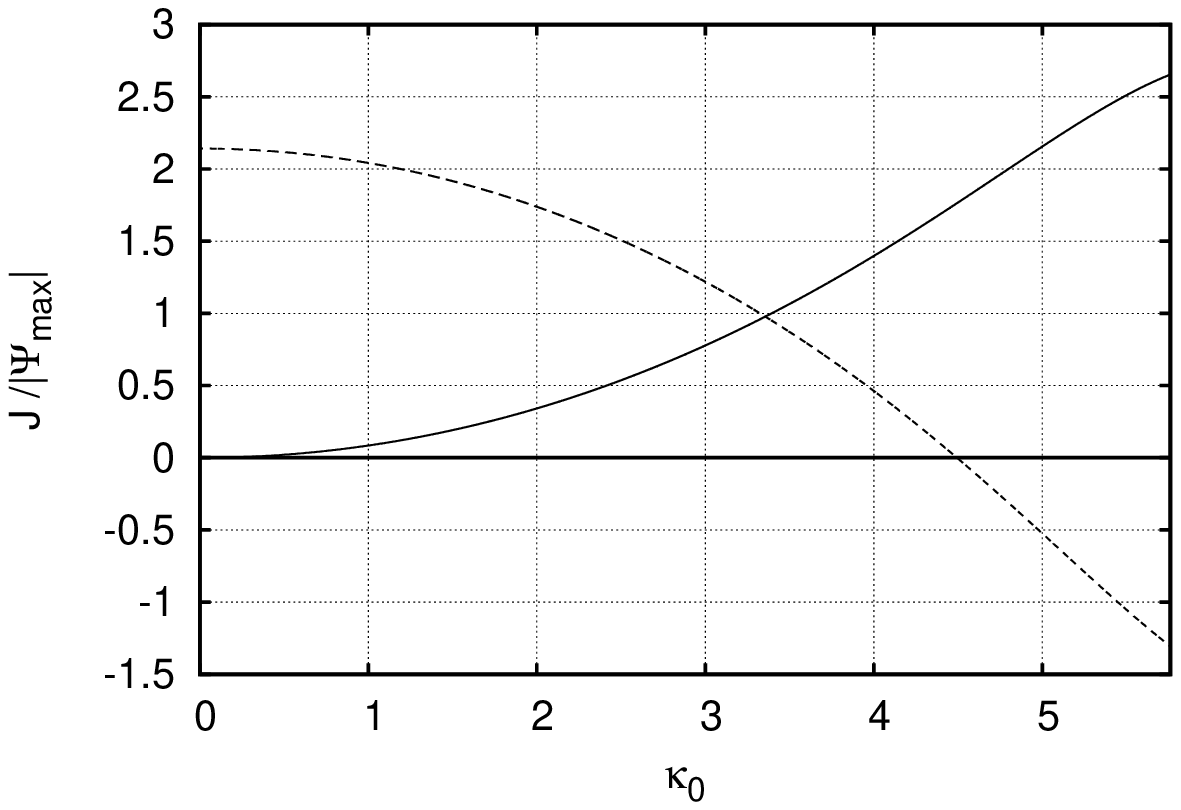}
 \caption{The total currents normalized $|\Psi_{\max}|$ are plotted 
against $\kappa_0$. We make $J_{tot}$ positive in this figure.
The left panel: Each line denotes $J_{tot}$ (thick solid line),
$J_{sur}$ (thick dashed line), $J_{tot}^{(+)}$ (thin solid line)
and  $J_{tot}^{(-)}$ (thin dashed line) respectively.
The right panel: Each line denotes $J_{tot}^{\kappa}$ (solid line)
and $J_{tot}^{\mu}$ (dashed line).
}
\label{fig:j_closed}
\end{figure*}

As explained before, we can calculate
one eigenfunction just by choosing one value 
of $\kappa_0$. By changing the value
of $\kappa_0$ and calculating the corresponding
solution for $\Psi$, we have a series of 
solutions which are shown in 
Fig.\ref{fig:j_0_kappa} and
Fig.\ref{fig:j_j_max}. Fig.\ref{fig:j_closed} 
displays the various total currents $J_{tot}$.

Fig. \ref{fig:j_0_kappa} shows how $j_0$ 
behaves for different value of $\kappa_0$ 
(left) and how  ${\cal M}_p / {\cal M}$ 
behaves against the value of $\kappa_0$ (right).
We set the parameters satisfying the relation
$r_s^{(s)2} \hat{\mu}_0 \bar{\rho} = -1$.
As seen from these panels, the function 
$j_0 (\kappa_0)$ has two special solutions 
which contain no surface currents 
at $\kappa_0 \sim$ 5.76, and at $\kappa_0 \sim$ 
9.10. Furthermore, there appear two singularities 
along this curve at $\kappa_0 \sim$ 4.49, and 
at $\kappa_0 \sim$ 7.73. The sign of the 
magnetic flux function changes at the 
singularities. The value of 
${\cal M}_p / {\cal M}$ decreases 
as the value of $\kappa_0$ increases until it 
reaches the first solution without
a surface current at $\kappa_0 \sim 5.76$
and the value of the ${\cal M}_p / {\cal M}$ 
reaches its minimum of ${\cal M}_p / {\cal M} 
= 0.417$.  Hereafter we will call the eigenvalue
$\kappa_0$ of the first solution without a 
surface current as $\kappa_m^1$ and the eigenvalue
$\kappa_0$ of the second solution without a 
surface current as $\kappa_m^2$. We also denote values of 
$\kappa_0$ at the first and the second singular 
points along this curve as $\kappa_s^1$ and 
$\kappa_s^2$, respectively.

We will discuss the behaviors of $\Psi$ and 
$j_0$ only around $\kappa_m^1$ in this paper.
However we see almost the similar behaviors 
for $\kappa_0 \sim \kappa_m^2$, which one
can find in several papers 
(\citealt{Ioka_Sasaki_2004}, \citealt{Duez_Mathis_2010},
\citealt{Yoshida_Kiuchi_Shibata_2012}).

We can classify the eigen solutions into four types 
according to the behaviors of the current 
densities as follows:
Type (a) -- solutions with $0 < \kappa_0 < \kappa^1_s$, 
Type (b) -- solutions with $\kappa_0 \sim \kappa_s^1$,
Type (c) -- solutions with $\kappa_s^1 < \kappa_0 
< \kappa_m^1$, 
and
Type (d) -- solution at $\kappa = \kappa_m^1$.
As we can see in Fig. \ref{fig:j_0_kappa}, 
concerning the ratio ${\cal M}_p / {\cal M}$,
the following relations hold:
\begin{eqnarray}
   & & \frac{{\cal M}_p} {{\cal M}} 
      \ \ (\mbox{Type a}) 
         > \frac{{\cal M}_p} {{\cal M}}
      \ \ (\mbox{Type b}) \nonumber \\
  &>& \frac{{\cal M}_p} {{\cal M}} 
      \ \ (\mbox{Type c}) 
        > \frac{{\cal M}_p} {{\cal M}} 
      \ \ (\mbox{Type d}).
\end{eqnarray}

In Fig.\ref{fig:j_j_max}, the distributions 
of $j_\varphi$ normalized by the maximum 
value of the flux function $|\Psi_{max}|$
along the normalized radius on the equatorial plane are displayed. 
In this figure the interior toroidal current 
$j_\varphi$, the component of $j_\varphi$ due to the 
$\kappa$ term and the component of $j_\varphi$ 
due to the $\mu$ term are displayed. The values 
of ($\kappa_0$, ${\cal M}_p / {\cal M}$)
in the panels are (a) (3.0, 0.701), (b) ($\sim 4.49$, 0.501), (c) (5.0, 0,452), 
and (d) (5.76, 0.417). In Fig.\ref{fig:j_closed}, we make that 
the sign of $J_{tot}$ is always positive. In other words, 
we multiply $J_{tot} / |J_{tot}|$
in order to plot the distributions of $J_{tot}$.
In the left panel of Fig.\ref{fig:j_closed},
each line denotes $J_{tot}$ (thick solid line),
$J_{tot}^{(+)}$ (thin solid line),
$J_{sur}$ (thick dashed line) and
$J_{tot}^{(-)}$ (thin dashed line).
In the right one, we decompose the $J_{tot}^{(+)}$
into the total $\kappa$ current and the total $\mu$ current.
The solid line show the $J_{tot}^\kappa$ and 
the dashed line show the $J_{tot}^{\mu}$ respectively.

It is remarkable that these models have strong 
toroidal magnetic fields contrary to 
configurations whose magnetic fields without
surface currents extend to the infinity, 
i.e. open fields, and cannot 
become large. From the panels in 
Fig.\ref{fig:j_j_max}, the configurations of 
Types (a) and (b) have the positive surface 
current in the $\varphi$ direction which is 
opposite to the interior negative 
current density in the $\varphi$ direction, 
while the configurations of Type (c) have 
the negative $\varphi$-surface current
while the interior $\varphi$-currents
are positive in almost all of the interior 
region. The model of Type (d) has no surface 
current. However, the $\varphi$ component of 
the interior currents are negative in the 
finite surface region while the $\varphi$ 
component of the interior currents in the
most inner region is positive.  
Therefore, in the range of $0 < \kappa_0 < \kappa_s^1$, 
$J_{tot}^{(+)} = J_{tot}^\mu + J_{tot}^\kappa$ and
in the range of $\kappa_s^1 < \kappa_0 \leq \kappa_m^1$, 
$J_{tot}^{(+)} = J_{tot}^\kappa$ and 
$J_{tot}^{(-)} = J_{tot}^\mu$ respectively.

The signs of the $\kappa$ term, the $\mu$ term 
and $j_\varphi$ of the solutions of Type (a) 
are all negative within the whole region as seen 
from Fig.\ref{fig:j_j_max}, while the surface 
current is positive because $j_0 > 0$.
Therefore, the strong toroidal magnetic fields 
of the solutions of Type (a) are sustained 
by the oppositely flowing surface current as 
has been explained for the reasons why the 
numerically exact open field configurations 
with surface currents can have large toroidal 
magnetic fields in the previous section. 
For Type (a) solutions, the toroidal magnetic 
field energy increases monotonously as the 
magnitude of the surface current becomes larger. 

The solutions of Type (b) show the extreme 
features corresponding to the singularity. 
Since the strength of the surface current 
becomes infinite, the $\kappa$ term becomes 
larger and the contribution from the $\mu$ term 
becomes nearly zero compared with that of 
the $\kappa$ term. Thus the non-force-free 
$\mu$ term contributes essentially nothing to 
the $\varphi$ component of the current density 
and so those solutions of Type (b) can be 
considered to be almost the same as those 
force-free configurations obtained by 
\cite{Broderick_Narayan_2008}.

As seen from these panels, the signs of the $\mu$ 
term and the $\kappa$ term are different from 
each other for configurations of Types (c) and (d).
In fact, the sign of the surface current changes 
from negative values to positive values 
between solutions of  Type (a) and those of Type 
(c). As a result, the interior currents can become 
negative near the stellar surface region. Since 
the surface currents are negative in this 
parameter range, the toroidal magnetic fields 
can be sustained by both the negative surface 
currents and the negative interior currents 
near the stellar surface region (see Fig.\ref{fig:j_closed}). 
The toroidal magnetic field energies become larger in this 
parameter range than those in the parameter range  
where only oppositely flowing surface currents
such as solutions of Types (a) and (b) 
are allowed. Moreover, the toroidal magnetic field energy
reaches its maximum value for the model 
(d) which has no surface current.
These phenomena imply that the effects of the
surface currents are not so large compared with 
those due to the interior currents near
the surface region which flow
oppositely to the interior currents further inside.

The model (d) has only the interior negative 
current region without surface currents 
in addition to the interior positive currents
in the inner part of the star. The effect
of the interior negative current is larger than 
that of configurations of Type (c) (see Fig.\ref{fig:j_closed}). 
The solutions obtained by \cite{Ioka_Sasaki_2004} and
\cite{Yoshida_Kiuchi_Shibata_2012} can be considered
to belong to the same type as the model (d)
except for the compressible densities.

As seen from the left panel of Fig.\ref{fig:j_closed}, 
the total current $J_{tot}$ becomes only 
slightly larger as $\kappa_0$ increases.
On the other hand, $J_{tot}^{(+)}$ increases 
rapidly beyond the $\kappa_0 = \kappa_s^1$ where 
$J_{tot}^{(-)}$ starts decreasing.  
This means that the negative current region 
can cancel much larger interior bulk positive 
current than the negative surface current.
Therefore, as we can see in Fig.\ref{fig:j_closed},  
the total $\kappa$ current becomes larger and 
the total $\mu$ current decreases rapidly
as $\kappa_0$ increases beyond the $\kappa_0 = \kappa_s^1$.
Since the ratio of ${\cal M}_p / {\cal M}$ reaches 
the minimum value for the model (d),
this kind of configuration without surface currents 
but with the negative interior current region has 
the strongest toroidal magnetic field energy 
among all the configurations as far as the
functional forms for $\kappa$ and $\mu$ are the
same as those chosen in this paper.

We have also calculated $N=1$ closed field 
configurations. Since we cannot integrate the 
source term analytically for compressible 
polytropes, we have used the shooting method 
to obtain the eigen solutions for the boundary
value problems (see \citealt{Ioka_Sasaki_2004}).
Obtained solutions of $N = 1$ polytropes have
the same tendencies as those for the $N=0$ 
solutions.  The ratio ${\cal M}_p / {\cal M}$ 
reaches its minimum value 0.349 
at $\kappa_m^1 =  7.42$. The toroidal 
magnetic field energy is slightly larger than 
that of the corresponding $N=0$ 
configuration.

\section{Discussion}

 \subsection{Open field configurations vs 
closed field configurations}

As we have calculated and seen in this paper,
the negative (oppositely flowing) surface 
currents and/or the negative (oppositely flowing)
interior currents seem to generate strong
toroidal magnetic fields within the stars.
We have obtained the configuration having 
the minimum value of ${\cal M}_p / {\cal M} 
\sim 0.697$ when $\hat{j}_0 = 7.5 \times 10^{-3}$ 
for the solutions of open fields
by taking the surface currents into account. 
On the other hand, the ratio 
${\cal M}_p / {\cal M}$ reaches 0.349
when all the magnetic field lines are closed
and confined inside the stellar surface for
$N=1$ polytropes. This value is much smaller 
than that for open field configurations with
surface currents. The functional forms for
the arbitrary functions and the boundary conditions 
of these two models, i.e. the closed filed  
solution and the open filed solution, are 
different from each other (see Eq.\ref{Eq:kappa}
and  Eq.\ref{Eq:jphi_closed}),
but both configurations contain the interior 
region where the interior currents are negative
(oppositely flowing) and there appear very strong 
toroidal magnetic fields.

From the right panels in Fig.\ref{Fig:j0_H_r_j_r},
the signs of the $\mu$ term and the $\kappa$ term 
are positive for our open field models
with negative surface currents. In other words,
the models do not contain the negative current 
regions. The open field models which we have 
obtained  correspond to models of the Type (a) 
for the closed field configurations.
Therefore, the stars could have stronger toroidal 
magnetic field energy if they can contain the 
negative interior current regions near the 
stellar surfaces. However, we could not find 
the functional form of $\kappa$ for which 
the negative interior current region appears 
as the  model of Type (d) in this paper.
In any case, the toroidal magnetic field energies 
of closed field models are larger than those of 
the corresponding open field models with negative
surface currents. It should be noted that
the model of Type (d) sustains the largest 
toroidal magnetic field energy among all of our 
solutions obtained in this paper.  We can 
conclude that the magnetized  equilibria with 
strong toroidal magnetic field energies 
would be the closed field configurations.

\subsection{Effects of compressibility to
toroidal magnetic fields}
\label{Sec:EOS}

We have employed polytropes with $N=1$ and 
$N=0$ in this paper.
Since we are mainly interested in the effects 
of surface currents, we adopt polytropes 
as equations of state, although polytropes
are too simple equations of state.
The different equations of state result in 
different density distributions as well as
different magnetic field structures, 
because the current density formula contains 
the density depending term (Eq.\ref{Eq:current}).
As we have described in 
Sec.\ref{Sec:closed_fields_model}, configurations for polytropes
with $N=1$ can sustain slightly larger toroidal 
magnetic fields than those with $N=0$, i.e. 
the minimum value of ${\cal M}_p / {\cal M}$
is $0.417$ for the $N=0$ polytropes  and
the minimum value of ${\cal M}_p / {\cal M} $ is 
$0.349$ for  $N=1$ polytropes as far as the other
parameters are the same. Therefore, 
the configurations with softer equations 
of state can sustain stronger toroidal magnetic 
fields for polytropes. 

How about for more realistic equations of 
state discussed by other authors ?
\cite{Kiuchi_Kotake_2008} calculated 
twisted-torus magnetized equilibrium states 
using some realistic equations of state at
zero temperature. Their method 
is the same as our method. Fig. 4 - Fig. 7 in 
their paper show the density contours and 
the magnetic field contours for different 
equations of state. The structures of the 
poloidal magnetic field lines and the regions 
of the toroidal magnetic fields within the
stellar surfaces are different among the 
different equations of state. For example, 
among models of the Shen's equation of state (EOS), 
the position where the toroidal magnetic
field attains its maximum strength is located 
near the stellar surface and the width of
the region where the toroidal magnetic fields
appear is relatively small, but for models with
the FPS's EOS, the position of the
maximum toroidal magnetic field is shifted to
the inner part of the star and the size of the
toroidal magnetic field region is 
much bigger than that for the Shen's EOS.
Although they did not calculated the toroidal 
magnetic field energies and the ratio 
${\cal M}_p / {\cal M}$, they showed
the ratio of the local strength of the toroidal 
magnetic field to the poloidal magnetic field 
($h$ in Table 4). The value of $h$ for FPS's
EOS is about as twice as that for  Shen's EOS. 
Therefore, although the influence of the equation 
of state might become more important if we would 
consider structures of neutron stars, it would not 
change the basic properties discussed in this paper
dramatically, although the values and/or the regions
for the toroidal magnetic fields would surely be
somewhat different from those obtained in this 
paper.

\subsection{Stability of configurations with 
oppositely flowing $\varphi$-currents 
within and/or on the stellar surfaces}
\label{Sec:stability}

It is very interesting and important to analyze
stability of our models for open magnetic fields
with surface currents. Since some of our 
solutions satisfy the Braithwaite's stability 
criteria,  Eq.(\ref{Eq:criteria}), our models 
could be stable. Although it is very difficult to
tell the stability for a certain model exactly,
we will be able to check the stability by
several non-exact ways and get rough idea
about the stability of the configuration.

First of all, we consider the stabilities of the 
magnetized stars with pure surface currents
and with no interior currents (see the left panel 
in Fig. \ref{Fig:dipole_quadrupole}).
The stability of the magnetized stars with 
surface currents in the surface region of
an infinitely thin width could be considered 
to be essentially the same as that of 
configurations with pure surface currents.
If the magnetized stars possess only the surface 
currents which generate the pure dipole magnetic 
fields outside the stars, their interior magnetic 
fields are uniform along the $z$ axis (see the 
left panel of Fig.\ref{Fig:dipole_quadrupole}). 
The magnetic fields of this kind of configuration
are unstable and decay within a few Alfv\'en time, 
because there is no toroidal magnetic fields
(\citealt{Markey_Tayler_1973}). 
As \cite{Flowers_Ruderman_1977} also explained 
the instability of this kind of configuration 
and \cite{Braithwaite_Spruit_2006}
carried out non-linear evolution of the 
instability by numerical simulations
and showed unstable growth of the initial 
stationary states as explained above. 
Therefore, the fields of the magnetized stars 
with pure surface currents can be considered 
to be unstable. 

By contrast, as for the configurations with 
surface currents, which might lay in,
for example, the crusts of the neutron stars, 
the magnetic fields could become stable.
In such configurations, we can assume that 
the widths of surface current layers are not 
infinitely thin any more and the finite 
Lorentz force acts on the surface currents. 
\cite{Flowers_Ruderman_1977} considered 
configurations with surface current layers as well as
with uniform magnetic fields and 
dipole magnetic fields inside and outside
of the stellar surface, respectively, and found
that those configurations with current layers
might be stable. In realistic situations as 
neutron stars, when the solid crusts of 
neutron stars form after their proto-neutron 
phase, the crusts could sustain the Lorenz 
force to themselves and they could prevent 
growth of the instability of magnetic fields.
For such situations, the magnetic fields can 
survive in much longer time than 
the Alfv\'en timescale.

Concerning direct computations of the evolutions
starting from the perturbed initial stationary 
states, \cite{Braithwaite_Spruit_2006} carried out
numerical evolutions of the  twisted-torus 
interior magnetic fields with solid crusts.
They included surface current layers with finite 
widths as their boundary condition for the 
magnetar's crust and used one of their 
quasi-stationary twisted configurations which 
they had obtained after long time simulations
as initial values. Their numerical model is 
similar to our solution with surface currents.
The magnetic fields of such stars do not 
decay within the Alfv\'en time scale in their 
simulations as far as the crusts can sustain 
the Lorentz force. Therefore, our twisted-torus
models with surface currents would be also 
stable configurations. 

Evolutions and stabilities of configurations
for closed magnetic fields were argued 
by \cite{Duez_Braithwaite_Mathis_2010}.
They performed numerical simulations using 
\cite{Duez_Mathis_2010} solutions as their initial states. 
They concluded that models with closed fields both with 
poloidal and toroidal magnetic fields do not 
show any sign of becoming unstable within 
their simulation time if the initial model
satisfy the stability criteria in Eq. (\ref{Eq:criteria}).
Therefore, the closed magnetic field models which are obtained in 
this paper would be stable configurations.

\subsection{Application to magnetars}
\label{Sec:magnetar}

It is important to find out natural mechanisms
to generate surface currents and/or their
origins if we apply our models with surface 
currents to real bodies such as magnetized 
neutron stars, especially to magnetars.
Magnetars are young neutron stars with very 
strong magnetic fields. The magnetars are
considered as source objects of special high 
energetic phenomena such as the anomalous X-rays 
emission and the soft gamma-ray emission. Thus
those pulsars are called the anomalous 
X-ray pulsars (AXPs) and the soft gamma-ray 
repeaters (SGRs), In particular, their 
high X-ray luminosities and giant flares 
have been considered to be deeply related 
to the strong magnetic fields of the stars 
(\citealt{Thompson_Duncan_1995, 
Thompson_Duncan_2001}). The magnetic fields 
are nearly dipole poloidal fields globally,
but there would be higher order (such as 
quadrupole and octopole) poloidal magnetic 
fields near the surface or toroidal fields 
winded up by rapid differential rotation 
during the proto-neutron star stages
(\citealt{Duncan_Thompson_1992}) inside 
the star. Before we apply our models with surface 
currents to the magnetars with strong toroidal 
magnetic fields, we need to clarify or
at least have some ideas about origins or
formation mechanisms for the oppositely
flowing surface currents or the discontinuity 
of the magnetic fields on the 
stellar surfaces. Then, what is the 
origin of the negative surface currents or
the negative current region? There might be 
two possibilities to explain it.
One is related to the crusts of neutron stars 
and the other is related to the magnetospheres 
around neutron stars.

Since the physics of the crusts of the neutron 
stars is too complicated and difficult to deal 
with, we only assume that the crusts consist of 
highly conductive solid matter. If the crusts are
highly conductive, the electric currents can 
exist within the crust regions. Then, the crusts 
can make parallel components of 
magnetic fields discontinuous near the stellar 
surfaces by the toroidal currents inside the crusts.
The magnetic fields are frozen to the matter 
and fixed to the crusts because of their high 
conductivity. On the other hand, the interior 
matter of the magnetars is not solid. Thus
the matter inside of the crusts can move 
differently from the crusts and the discontinuities 
of the magnetic fields would be born between 
the crusts and the interior regions. 
The interior fields begin to spread toward 
the stellar surfaces by the some kind of 
magnetic diffusion (\citealt{Braithwaite_Spruit_2006})
the discontinuities would be enlarged by the 
magnetic pressure. As we have seen before,  
\cite{Braithwaite_Spruit_2006} simulated this 
kind of configuration and found the growing of 
the Lorentz stress in the crusts.
From the direction of the discontinuity,
we expect that the stress is tensile one globally.
If the crusts are cracked by the stress, 
it would result in flares of SGRs. Following 
this scenario, our models with strong toroidal 
fields as well as surface currents are 
considered as stationary states of the crusts
with strong Lorentz forces before occurrence of
giant flares. If a part of the crusts is cracking,
the magnetic energy and the helicity are injected 
from the stars and would produce magnetized flows
(\citealt{Takahashi_Asano_Matsumoto_2009, Takahashi_Asano_Matsumoto_2011},
\citealt{Matsumoto_Masada_Asano_Shibata_2011}).
These kinds of magnetized outbursts would be giant 
flares of SGRs. We will consider this process by 
using our  models with surface currents in the
following. 

At first, the surface currents in the crusts
can sustain the strong toroidal fields by  
bending the poloidal magnetic fields as 
shown by the model with  
$\hat{j}_0 = 7.5 \times 10^{-3}$ in the right panel 
of Fig.\ref{Fig:u_flux}. When the Lorentz force 
exceeds a certain critical value, a part of the 
crust begins to crack. We can consider this phenomena 
as decreasing the strength of the surface current, 
because a part of the conductive matter is 
disturbed by the cracking. We assume that a certain 
cracking reduces the value of $\hat{j}_0$ 
from $7.5 \times10^{-3}$ to $2.5 \times10^{-3}$ 
as an extreme example.
The surface current with $\hat{j}_0 = 2.5 \times 10^{-3}$
cannot sustain the toroidal magnetic fields any
more which the surface current with 
$\hat{j}_0 = 7.5 \times 10^{-3}$ has sustained. 
The toroidal magnetic energy and/or the
magnetic helicity would be transferred  out into 
the outside of the star in order to relax 
to the stationary state with 
$\hat{j}_0 = 2.5 \times 10^{-3}$ 
(transition from the right panel to the left panel 
in Fig.\ref{Fig:u_flux}). Through this cracking,
the dimensionless toroidal magnetic field  
energy ${\cal \hat{M}}_t$ 
changes from $2.49 \times 10^{-2}$ to $9.92 
\times 10^{-3}$ according to our calculations. 
Therefore, about 60 \% of the toroidal magnetic 
field energy would be released during the cracking.
Although this is an extreme example,
it is natural that the injection of the 
magnetic helicity and 
the release of the toroidal field come 
from the transition of the magnetized equilibria 
by the phenomena such as cracks of the crusts 
which reduce the surface current strength.

Another possibility is the effect due to the
magnetosphere which excites oppositely
flowing current densities near the stellar 
surfaces. \cite{Colaiuda_et_al_2008} discussed the 
importance of the magnetosphere as the boundary 
conditions for both the poloidal and toroidal 
magnetic fields on the stellar surfaces.
Our present models and many configurations in
other previous works assume that the outside 
of the star is vacuum where the current density 
and the toroidal magnetic fields do not exit.
However, the presence of the magnetosphere 
changes the boundary condition for 
the magnetic field. This change of the
boundary condition would significantly 
influence on the magnetic field configurations 
as we have seen. 

As for rotation powered pulsars, there have
been many investigations about their magnetospheric
phenomena such as pulsar winds (see e.g. 
\citealt{Goldreich_Julian_1969}).
Their rapid rotations produce enormously
large electrical forces and the surface 
charged layers could not stay in their stationary
states. Charged particles run away from the 
stellar surfaces and form the pulsar magnetospheres
(see, for the recent particle pulsar wind 
simulations, 
\citealt{Wada_Shibata_2007, Wada_Shibata_2011}). 
This charged particles  would produce the strong 
currents outside the star and the twisted 
magnetosphere would form.

The rotational speeds and the strengths of the 
magnetic fields of the magnetars are 
different from those of pulsars, but there 
would be some kinds of magnetospheres around 
the magnetars. The recent X-ray spectral 
observations show the presence of a 
magnetosphere for the magnetar 
(\citealt{Rea_et_al_2009}).
Very recently, \cite{Viagno_Pons_Miralles_2011} 
computed numerically a force-free twisted 
magnetar magnetosphere. They treated the
stationary state of the magnetized star 
as an inner boundary condition
for the magnetosphere. We can see the 
various magnetospheric structures by applying
different inner boundary conditions in their paper.
If the magnetosphere forms the oppositely
flowing  toroidal currents near the stellar 
surface, the magnetized star can sustain the 
strong toroidal magnetic field energy inside the 
star. However, the details of the calculation are 
beyond the scope of the present study.

\section{Summary and Conclusion}

In this paper we have dealt with the effects of 
surface currents upon the interior toroidal 
magnetic fields. We have shown that the 
oppositely flowing surface currents 
can sustain the strong toroidal magnetic field 
energy inside the star both for the 
open and closed field configurations.

In the open field models, we have found that 
there is an upper bound of the total current 
of the star for a fixed set of parameter
values. Increasing the maximum strength of the
toroidal magnetic field decreases the region 
of the toroidal magnetic field due to this 
upper bound. Therefore this upper bound limits 
the ratio of the poloidal magnetic field energy 
to the total magnetic field energy.
To exceed this upper bound, magnetized star 
needs the oppositely flowing surface 
currents to the interior toroidal currents.
The interior current can overcome the upper bound
and the ratio ${\cal M}_p /{\cal M}$
decreases significantly, because 
the surface current counteracts the interior 
toroidal current.
 
In the closed field models, we have found that 
a model with an oppositely flowing current region but
with no surface current can sustain the 
strongest toroidal magnetic field among all 
of our magnetized stationary states.
The negative surface currents can sustain the 
strong toroidal magnetic fields in the models 
with closed magnetic fields.
However, the strengths of the toroidal magnetic 
fields for models with negative surface currents
cannot exceed a critical value even if the 
strength of the surface current becomes infinity.
In order to overcome the critical value, 
the negative current region is required.
Increasing the size of the negative interior
current region decreases the negative surface 
current. As a result, the toroidal magnetic 
fields become the strongest when not only 
the negative interior current region becomes 
the largest but also the surface current 
disappears.

It should be also noted that, although we have not
imposed a condition 
$d \Psi/d r|_{r = r_s^{(s)}}= 0$,
the obtained two eigenfunctions without 
surface currents fulfill this condition and,
moreover, that the values of 
${\cal M}_p/{\cal M}$ are very small.
This implies that by computing a series of
eigen states with surface currents as well as
with oppositely flowing interior currents we
could have easily reached an eigen state whose
eigen function behaves very {\it smoothly}
for which the role of the toroidal magnetic fields 
becomes very important.  Furthermore, 
it is remarkable that these 
solutions obtained by considering {\it in the 
wider functional space} without no
restrictions about the slopes of
the functions correspond to those solutions 
obtained by other authors 
(\citealt{Duez_Mathis_2010}). 

We have applied the models of open magnetic fields 
with surface currents to explain the strong 
hidden toroidal magnetic fields inside the 
magnetars. We have considered two possibilities  
as the origin of their surface currents. 
One possibility is related to the crusts of 
the magnetars. Since the crusts are made of 
the solid matter, it could make the
magnetic fields discontinuous at the crusts and 
the surface currents would appear due to these 
discontinuities. The magnetized stars can 
sustain the strong toroidal magnetic field 
energy by bending the poloidal magnetic 
fields within the crust zones.

The other possibility for the excitement of
oppositely flowing currents inside and/or
on the stellar surface might be related to 
magnetospheres which produce the oppositely 
flowing toroidal currents near and/or on the 
stellar surface. This kind of magnetosphere 
would also sustain the strong toroidal magnetic 
field energy inside the star.
These models might be the key to reveal the 
mechanism of the giant flares of the magnetars.

\section*{ACKNOWLEDGEMENTS}

The authors would like to thank the anonymous reviewer for useful
comments and suggestions that helped us to improve this paper. 
K.F. would like to thank Dr. Braithwaite for his very 
exciting discussion and drinking during the "Magnetic Fields 
in the Universe III" conference in Zakopane, Poland.
K.F. would also like to thank the members of the Plasma physics 
seminar at NAOJ for their very interesting discussion.
This work was supported by Grant-in-Aid for JSPS Fellows.

\bibliographystyle{mn}


\appendix

\section{Physical quantities and 
accuracies of numerical solutions}

\subsection{Global physical quantities}
\label{App:Physics}

In order to see the global characteristics of 
magnetized equilibria, we define several global 
physical quantities as follows:
\begin{eqnarray}
   W \equiv \frac{1}{2}\int \rho \phi_g \, d^3 \Vec{r} \ ,
\end{eqnarray}
\begin{eqnarray}
   T \equiv \frac{1}{2} \int \rho (R \Omega)^2 \, d^3 \Vec{r} \ ,
\end{eqnarray}
\begin{eqnarray}
   \Pi \equiv \int p \, d^3 \Vec{r} \ ,
\end{eqnarray}
\begin{eqnarray}
   F \equiv \int r \cdot \left(\frac{\Vec{j}}{c} \times \Vec{B} \right) \, d^3 \Vec{r} \ ,
\end{eqnarray}
where, $W$, $T$, $\Pi$ and $F$ are the 
gravitational energy, the kinetic energy, 
the volume integral of the pressure and 
the volume integral of the quantity which
is an inner product between the 
position vector and the Lorentz force vector,
i.e. which is related to the work due to the
Lorentz force,  
respectively. We also define the magnetic field 
energies of the star as below:
\begin{eqnarray}
  {\cal M}_t = \frac{1}{8\pi} \int B_\varphi^2  \, dV ,
\end{eqnarray}
\begin{eqnarray}
 {\cal M}_p = \frac{1}{8\pi} \int (B_r^2 + B_\theta^2) \, dV 
\label{Eq:M_p}
\end{eqnarray}
\begin{eqnarray}
 {\cal M} = {\cal M}_p + {\cal M}_t .
\end{eqnarray}
Here, ${\cal M}_p$, ${\cal M}_t$ and ${\cal M}$ 
are the poloidal magnetic field energy, 
the toroidal magnetic field energy and 
the total magnetic field energy, respectively. 
We use these values to evaluate the ratio 
of the ${\cal M}_p / {\cal M}$.

\subsection{Dimensionless physical quantities}
\label{App:Dimensionless}

We display dimensionless forms of  other 
physical quantities as below:
\begin{eqnarray}
 \hat{r} \equiv \frac{r}{r_e}
\end{eqnarray}
\begin{eqnarray}
 \hat{\rho} \equiv \frac{\rho}{\rho_{\max}}
\end{eqnarray}
\begin{eqnarray}
   \hat{\phi}_g \equiv \frac{\phi_g}{4 \pi G r_e^2 \rho_{\mathrm{max}}} \ ,
\end{eqnarray}
\begin{eqnarray}
   \hat{\Omega} \equiv \frac{\Omega}{\sqrt{{4 \pi G \rho_{\mathrm{max}}}}} \ ,
\end{eqnarray}
\begin{eqnarray}
 \hat{\kappa} \equiv \frac{\kappa}{\sqrt{4\pi G} r_e^2 \rho_{\max}} \ ,
\end{eqnarray}
\begin{eqnarray}
 \hat{\mu} \equiv \frac{\mu}{\sqrt{4 \pi G}/r_e} \ ,
\end{eqnarray}
\begin{eqnarray}
   \hat{A}_\varphi \equiv \frac{A_\varphi}{\sqrt{4 \pi G }r_e^2 \rho_{\max} } \ ,
\end{eqnarray}
\begin{eqnarray}
 \hat{\Psi} \equiv \frac{\Psi}{\sqrt{4 \pi G }r_e^3 \rho_{\max} } \ ,
\end{eqnarray}
\begin{eqnarray}
   \hat{j}_\varphi \equiv \frac{j_\varphi}{\sqrt{4\pi G}\rho_{\max} c} \ ,
\end{eqnarray}
where $G$, $\rho_{\rm max}$ and $r_e$ are the 
gravitational constant, the maximum value of 
the density and the equatorial radius, 
respectively.

\subsection{Setting of the mesh points 
and accuracies of numerical solutions}
\label{App:VC}

 \begin{figure}
    \begin{center}
     \includegraphics[scale=0.6]{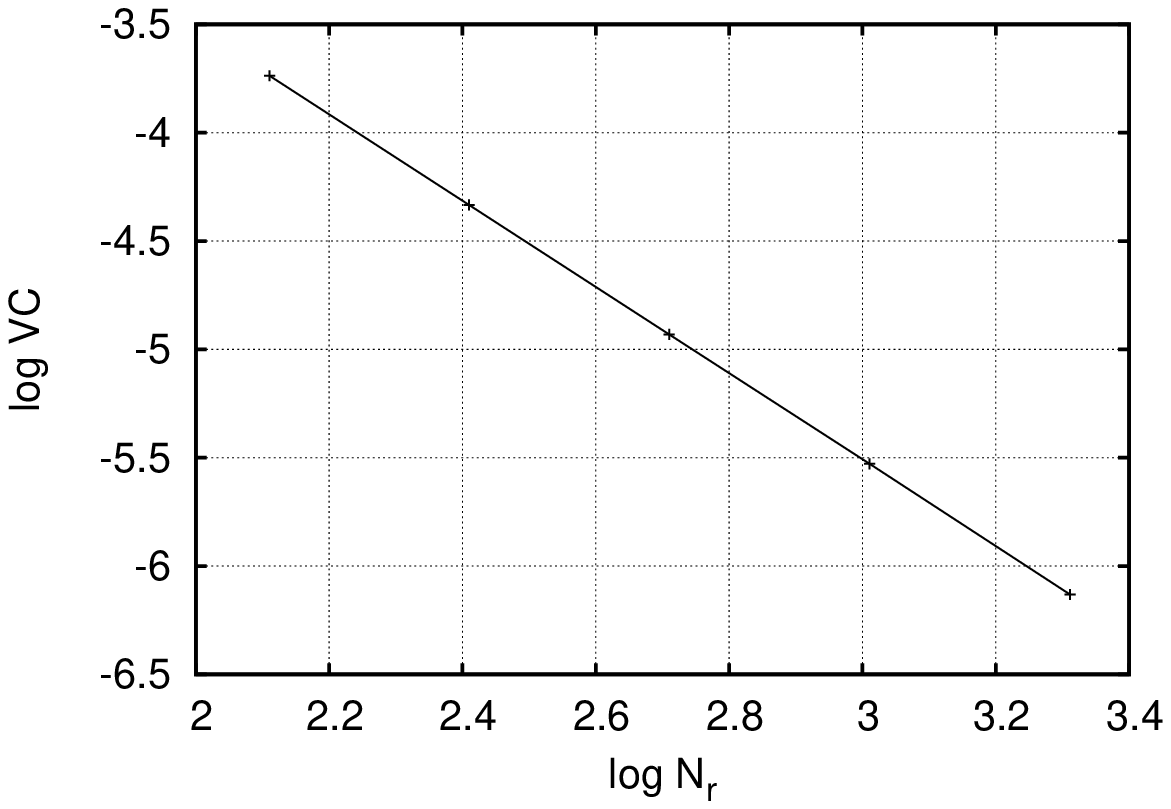}
    \end{center}
    \caption{The virial quantity VC is plotted 
    against the number of the grid point in 
    the $r$-direction ($N_r$)}
    \label{Fig:VC}
 \end{figure}
 
In order to examine the convergences and 
the accuracies of our numerical solutions,  
we use the virial relation as follows:
\begin{eqnarray}
   \mathrm{VC} \equiv \frac{|2T + W + 3\Pi + F|}{|W|}.
\end{eqnarray}
Since this value must vanish for exact 
equilibrium solutions, we can estimate the 
global numerical errors by this quantity.
This value depends on the mesh size, 
because we use finite difference method by 
discretizing the computational region into 
finite meshes. We use two computational regions 
because we need to integrate Eq. (\ref{Eq:M_p}) 
even at a far distant place from the star. One 
is  $\hat{r} = [0,1]$, $\theta = [0,\pi]$, i.e.
for the stellar interior region, and 
the other is $\hat{r} = [1,100]$, 
$\theta = [0, \pi]$, i.e. for the vacuum region.
We discretize the stellar interior region 
into mesh points with an equal interval 
$d\hat{r}$ and the vacuum region into mesh
points with an equal interval $d\hat{s}$ where 
$\hat{s}$ is defined as below:
\begin{eqnarray}
 \hat{s} = \frac{\hat{r}}{1 + \hat{r}}.
\end{eqnarray}
The region of $\hat{r}[1:100]$ corresponds 
to the region of $\hat{s}[1/2:100/101]$. 
We use the same mesh number 
in these two regions. As for the $\theta$-direction, 
we discretize it into mesh points with an 
equal interval  ($d \theta$).
We compute some configurations and change the 
number of grid points 
in the $r$-direction $N_r$ and fixed the 
number of 
grid points in the $\theta$-direction, 
$N_\theta = 513$.  
Fig.\ref{Fig:VC} shows 
the convergence of the VC with increasing 
$N_r$.
We see almost the 2nd order accuracies 
from the convergent tendency of solutions 
from Fig. \ref{Fig:VC}
because we use the 2nd order integral scheme,
i.e. Simpson's scheme.
If we use mesh numbers, $N_r = 513$ and
$N_\theta = 513$, the typical VC 
values are smaller than $1.0 \times 10^{-4} 
\sim 10^{-5}$. These values are small 
enough to be able to consider the 
systems are in equilibrium states 
(see also \citealt{Hachisu_1986a}). 
Thus we fixed the mesh numbers $N_r = 513$ 
and $N_\theta = 513$ 
during all calculations in this paper 
to obtain accurate magnetized configurations.

\section{Explicit forms of $K_1$ and 
\lowercase{$j_0$}
for configurations with surface currents of 
the dipole type}
\label{App:K_1}

From the Eq.(\ref{Eq:K_1}), we obtain the explicit 
form of $K_1$ as
\begin{eqnarray}
  K_1 &= 4 \pi \mu_0 \bar{\rho} \frac{r_s^{(s)2} }{ \kappa_0^4}
\Bigg(
\frac{-\cos \kappa_0 - \kappa_0 \sin \kappa_0 }{ \sin \kappa_0 - \kappa_0 \cos \kappa_0}
\Bigg) 
\nonumber \\
& \Bigg( (3 - \kappa_0^2) \sin(\kappa_0) - 3 \kappa_0 \cos \kappa_0 \Bigg).
\end{eqnarray}
The coefficient $j_0$ for the dipole type
distribution of the surface current 
treated in this paper can be
expressed  as follows:
\begin{eqnarray}
\begin{split}
&  j_0 \equiv  \frac{1}{4 \pi r \sin^2 \theta}\P{\Psi (r, \theta)}{r}\Bigg|_{r=r_s^{(s)}} \\
& = \frac{1}{4\pi r_s^{(s)}}\Bigg [
 \frac{K_1}{\kappa_0} \Big\{  (\kappa_0^2 -1 ) \sin \kappa_0 + \kappa_0 \cos\kappa_0 \Big\} \\
& + \frac{4\pi \mu_0 \bar{\rho} r_s^{(s)2}}{\kappa_0^5} \Big\{ (2 \kappa_0^3 - 3 \kappa_0) \sin^2 \kappa_0 \\
&+ (-2\kappa_0^4 + 8 \kappa_0^2 -3) \cos \kappa_0 \sin \kappa_0 
 + ( 3 \kappa_0 - 4 \kappa_0^3) \cos^2 \kappa_0
\Big\} \\
& + \frac{4\pi \mu_0 \bar{\rho} r_s^{(s)2}}{\kappa_0^3} \Big\{ - \kappa_0 \sin^2 \kappa_0 
 + (\kappa_0^2 - 1) \cos \kappa_0 \sin \kappa_0 + \kappa_0  \cos^2 \kappa_0
\Big\}
 \Bigg ] .
\end{split}
\end{eqnarray}

\end{document}